\documentclass[10pt,onecolumn,aps,prd,preprintnumbers,showpacs,superscriptaddress,nofootinbib,amsmath,amssymb,floats,floatfix,showkeys,notitlepage,longbibliography]{revtex4-2}
\usepackage{orcidlink}
\usepackage{comment}
\usepackage{lipsum}
\usepackage{graphicx}
\usepackage{subfigure}
\usepackage{palatino}
\usepackage{sans}
\usepackage{hyperref}
\hypersetup{colorlinks=true,linkcolor=blue,urlcolor=blue,citecolor=blue}
\usepackage[toc,page]{appendix}
\usepackage[normalem]{ulem}
\usepackage{adjustbox}
\usepackage{latexsym}
\usepackage{amsmath}
\numberwithin{equation}{section}
\usepackage{amssymb}
\usepackage{amsfonts}
\usepackage{dcolumn}
\usepackage{bm}
\usepackage{tikz}
\usetikzlibrary{decorations.pathmorphing}
\usepackage{bigints}
\usepackage{array,tabularx,multirow,booktabs}
\usepackage[tracking=true]{microtype}
\usepackage{soul} 
\SetTracking{}{500}
\SetTracking{encoding={*}, shape=sc}{40}
\UseRawInputEncoding 
\allowdisplaybreaks

\usepackage[utf8]{inputenc}
\usepackage{algorithm}
\usepackage{algorithmicx}
\usepackage{algpseudocode}
\usepackage{xcolor} 

\begin{document} \sloppy

\title{Finite Distance Corrections to Vacuum Birefringence in Strong Gravitational and Electromagnetic Fields}

\author{Ali \"Ovg\"un \orcidlink{0000-0002-9889-342X}}
\email{ali.ovgun@emu.edu.tr}
\affiliation{Physics Department, Eastern Mediterranean University, Famagusta, 99628 North
Cyprus via Mersin 10, Turkiye.}

\author{Reggie C. Pantig \orcidlink{0000-0002-3101-8591}} 
\email{rcpantig@mapua.edu.ph}
\affiliation{Physics Department, School of Foundational Studies and Education, Map\'ua University, 658 Muralla St., Intramuros, Manila 1002, Philippines.}

\begin{abstract}
We study polarization dependent photon propagation in static, spherically symmetric spacetimes permeated by strong magnetic fields, with the aim of quantifying how finite emission and detection radii modify vacuum birefringence signals. Working in the geometric optics limit of NLED, we formulate the two polarization modes as null geodesics of distinct effective (optical) metrics. We then develop a controlled weak-coupling expansion that cleanly separates the standard gravitational deflection from the birefringent contribution induced by the electromagnetic nonlinearity. Using a finite distance Gauss-Bonnet construction on the associated optical manifolds, we derive a general expression for the \emph{differential} bending angle in which the source and observer are kept at arbitrary radii, thereby extending the usual scattering-at-infinity treatment. As benchmark applications, we specialize our results to the Euler-Heisenberg effective action of quantum electrodynamics (QED) and to Born-Infeld electrodynamics. We find that the observable birefringence is generically reduced by finite-distance truncation of the curvature flux, and we provide explicit correction series suitable for data analysis. For magnetar-motivated dipole-like falloff, the same geometric truncation can reduce the predicted polarization-dependent
deflection at the order-tens-of-percent level for surface/near-surface emission; in a simple one-sided (outward-only) benchmark
the suppression can approach $\sim 1/2$ for near-limb rays. Since realistic dipole magnetospheres are axisymmetric rather than
spherically symmetric, we present this as an illustrative scaling estimate and leave a fully axisymmetric treatment to future work. Our results furnish a necessary finite-distance calibration for interpreting current and future X-ray polarimetry measurements and for placing unbiased constraints on strong-field QED and broader NLED parameters.
\end{abstract}


\maketitle

\section{Introduction}\label{sec1}

The interaction of light with strong gravitational and electromagnetic fields provides a uniquely clean arena in which fundamental physics becomes observationally testable. In classical general relativity, light propagation is achromatic and polarization blind: null rays follow the spacetime light cone, and gravitational lensing is a purely geometric effect \cite{Einstein_2005,Dyson:1920cwa}. Quantum field theory in external backgrounds, however, predicts that the vacuum itself acquires nontrivial optical properties in sufficiently intense electromagnetic fields. In particular, vacuum polarization in QED leads to the celebrated Euler-Heisenberg effective action and to birefringence and dispersion of photon modes in strong magnetic fields \cite{Heisenberg:1936nmg,Bialynicka-Birula:1970nlh,Adler:1971wn,Schwinger:1951nm}. Beyond QED, NLED also arises as an effective description in broader contexts, including string-inspired constructions, and admits a rich phenomenology when coupled to gravity \cite{Fradkin:1985qd,Tseytlin:1995uq,Sorokin:2021tge,Kruglov:2015yua,Capozziello:2025wwl,DeBianchi:2025bgn,Capozziello:2024ucm}.

The advent of horizon-scale imaging by the Event Horizon Telescope (EHT), exemplified by the image of Sagittarius A*, has opened a new precision window for testing gravity and fundamental physics in the strong-field regime, where deviations from the Kerr paradigm can be constrained directly at the scale of the photon ring \cite{Vagnozzi:2022moj}. In this context, NLED provides a particularly well-motivated arena: magnetically charged black holes sourced by NLED can yield shadow and lensing signatures that are, in principle, accessible to EHT analyses and can be used to bound the underlying electromagnetic nonlinearity \cite{Allahyari:2019jqz,Chen:2023trn}. More broadly, combining electromagnetic probes (shadow, lensing) with dynamical and particle-physics observables (quasinormal modes, greybody bounds, and propagation effects) offers a multi-messenger strategy to break degeneracies among model parameters \cite{Pantig:2022gih,Lambiase:2005gt}. Recent studies of rotating Einstein-Euler-Heisenberg black holes further demonstrate how QED-inspired nonlinearities can imprint simultaneously on shadows and ringdown spectra, reinforcing the relevance of strong-field birefringent and dispersive effects for upcoming high-resolution observations \cite{Lambiase:2024lvo}. NLED provides a well-motivated extension of Maxwell theory that captures high-field corrections arising both as effective quantum phenomena, such as the Euler-Heisenberg vacuum polarization effects underlying photon dispersion and photon splitting in strong backgrounds \cite{Bialynicka-Birula:1970nlh,Adler:1971wn}, and as fundamental structures emerging in broader frameworks, including string-inspired constructions \cite{Fradkin:1985qd,Tseytlin:1995uq}. When coupled to general relativity, NLED generically alters the near-horizon and strong-curvature regimes and has long been known to admit black-hole geometries whose properties can deviate markedly from the Reissner-Nordstr\"om solution \cite{Pellicer:1969cf,deOliveira:1994in,Yajima:2000kw,Ruffini:2013hia,Kruglov:2015yua}. A particularly influential development is that suitable NLED sources can yield regular (nonsingular) black holes, providing controlled arenas to explore the interplay between energy conditions, effective stress tensors, and astrophysical observables \cite{Ayon-Beato:1998hmi,Bronnikov:2000vy,Dymnikova:2004zc,Balart:2014cga,Fan:2016hvf,Bambi:2013ufa,Li:2024rbw}. These geometries have renewed relevance because NLED-induced modifications can imprint themselves on wave and ray probes (e.g., quasinormal spectra, birefringence, and optical signatures such as weak deflection and shadow deformations \cite{Breton:2021mju,Ovgun:2019wej,Okyay:2021nnh}). In this context, the recent construction of new black hole solutions within both second- and first-order formulations of NLED sharpens the theoretical landscape by clarifying how distinct variational principles and constitutive structures map into physically inequivalent spacetime geometries, thereby enlarging the space of consistent strong-field models testable with gravitational and electromagnetic observations \cite{Verbin:2024ewl}, and complementing related strong-field extensions in modified gravity settings, including scalarized dyonic configurations \cite{Brihaye:2021ich}.

These effects are not merely academic: magnetars, with surface magnetic fields of order $10^{14}$-$10^{15}\,$G, constitute natural laboratories where QED vacuum polarization is expected to leave measurable imprints on X-ray propagation and polarization \cite{Duncan:1992hi,Kaspi:2017fwg,Mignani:2016fwz,Taverna:2022jgl}. The emergence of dedicated X-ray polarimetry missions has therefore transformed vacuum birefringence from a theoretical hallmark into a precision diagnostic, sharpening the need for theoretical predictions that correctly encode the relevant geometry of emission and detection \cite{Weisskopf_2016,eXTP:2018anb,Taverna:2024uop,Lai:2022knd}. In realistic magnetar observations, photons are emitted at finite radius (often from the stellar surface or magnetosphere) and are detected at large but finite distance, so the accumulated polarization-dependent bending and phase shift can differ systematically from idealized scattering calculations.

A powerful way to formulate photon propagation in NLED is the effective (optical) metric approach. In the geometric-optics limit, the eikonal wave vector satisfies a Fresnel equation that generically factorizes into two polarization branches, implying that the two photon modes propagate along null geodesics of two distinct effective geometries \cite{Boillat:1970gw,Plebanski:1959ff,Novello:1999pg,DeLorenci:2000yh}. In this picture, birefringence becomes a manifestly geometric phenomenon: it is encoded in the inequivalence of the corresponding optical manifolds, rather than introduced by hand through refractive indices.
This viewpoint has been widely exploited to compute polarization-dependent deflection angles and related strong-field observables in birefringent NLED models (notably QED-type effective actions) under asymptotic assumptions \cite{Bialynicka-Birula:1970nlh,Adler:1971wn,Born:1933pep,Born:1934gh}. 
Exceptional NLEDs also exist: in particular, Born-Infeld electrodynamics is \emph{non-birefringent}--the two polarization modes propagate on the same effective light cone (up to conformal rescaling) \cite{Boillat:1970gw,Obukhov:2002xa}. NLED generically modifies light propagation because electromagnetic perturbations propagate on an effective (field-dependent) geometry, leading to nontrivial causal and optical phenomena so strong, in fact, that closed, space-like photon paths can arise in suitable NLED backgrounds \cite{Novello:2001fv}. A systematic way to characterize these effects is provided by Fresnel (birefringence) analysis, which derives the dispersion relations and polarization-dependent light cones for general NLED theories \cite{Obukhov:2002xa}. In gravitational settings, NLED also supports magnetically charged regular/“nonlinear magnetic” black holes whose dynamical viability can be tested via stability analyses \cite{Breton:2005ye} and by studying their response to electromagnetic perturbations and late-time relaxation (ringdown) \cite{Toshmatov:2018tyo,Toshmatov:2019gxg}. On the observational side, regular spacetimes such as Bardeen and Ayón–Beato–García exhibit distinctive circular-geodesic structure and disk phenomenology—impacting ISCO properties, profiled spectral lines from Keplerian disks/rings, and even producing “ghost” lensing images in no-horizon configurations \cite{Stuchlik:2014qja,Schee:2016mjd,Schee:2015nua,Schee:2019gki}.

Strong field gravitational lensing provides a clean probe of the near horizon geometry, producing a sequence of relativistic images whose positions and magnifications encode the properties of the underlying compact object. The foundational strong lensing framework for Schwarzschild spacetime and its relativistic images was developed in Refs.~\cite{Virbhadra:1999nm,Virbhadra:2008ws}. Extending these ideas to NLED, Born–Infeld black holes were shown to yield distinctive lensing signatures relative to Maxwell electrovacuum \cite{Eiroa:2005ag}, and more generally light propagation can, in principle, discriminate the effective “charge” content of NLED black holes through characteristic deviations in the deflection and optical observables \cite{Toshmatov:2021fgm}. More recently, combining strong lensing observables with horizon scale imaging has emerged as a powerful avenue to constrain modified gravity models using both lensing data and black-hole shadow measurements \cite{Kuang:2022ojj}. In particular, braneworld black holes exhibit characteristic modifications in their lensing and retrolensing observables compared to GR expectations, providing an astrophysical window into extra-dimensional corrections \cite{Abdujabbarov:2017pfw}. Realistic propagation effects can further reshape these signatures: the presence of plasma alters photon trajectories and observable deflection, and this impact has been quantified for Schwarzschild black holes embedded in perfect-fluid dark matter environments \cite{Atamurotov:2021hoq} as well as for Schwarzschild–MOG black holes, where both modified gravity and dispersive plasma effects jointly influence the weak-lensing regime \cite{Atamurotov:2021qds}. Regular/quantum-corrected black holes provide a well-motivated arena where both shadow and weak lensing observables can deviate from the Schwarzschild/Kerr expectations, especially once realistic dispersive media are included. In asymptotically safe gravity, plasma-induced refraction can significantly reshape the apparent shadow size and distort weak-deflection predictions for regular black holes, strengthening the case for joint shadow–lensing tests in non-vacuum environments \cite{Turakhonov:2025ojy}. Complementarily, charged black holes in regularized (4D) Einstein–Gauss–Bonnet gravity exhibit modified bending angles and image properties relative to GR, offering an additional probe of higher-curvature corrections through gravitational lensing \cite{Kumar:2020sag}. Beyond modified gravity, NLED models can generate regular (and even horizonless ultracompact) geometries with distinct photon-sphere structure and shadow imprints, making shadow measurements a sharp discriminator among NLED theories and compact-object classes \cite{KumarWalia:2024yxn}. Recent work has sharpened the consistency requirements behind “regular black holes sourced by nonlinear electromagnetic fields,” emphasizing that singularity resolution must be confronted simultaneously with the NLED field equations, global structure, and the effective (possibly birefringent) causal cones governing photon propagation \cite{Bokulic:2025brf,Babaei-Aghbolagh:2025tim}. In strictly stationary settings and in broader analyses of NLED-coupled spacetimes, nontrivial integrability and energy–momentum constraints can strongly restrict which NLED models can genuinely support nonsingular geometries \cite{Bokulic:2021xom,Bokulic:2022cyk}, complementing earlier discussions on construction subtleties and consistency checks for “regular BH recipes” \cite{Fan:2016hvf,Toshmatov:2018cks}. A central issue is stability: linear analyses for self-gravitating NLED black holes and magnetic configurations reveal nontrivial mode conditions \cite{Moreno:2002gg,Nomura:2020tpc}, and recent results indicate that broad classes of NLED nonsingular black holes can suffer instabilities, substantially constraining their viability as realistic endpoints of collapse \cite{DeFelice:2024seu}. These theoretical constraints are further tightened by laboratory/astrophysical bounds on nonlinear corrections to Maxwell theory, including limits from light-by-light (($\gamma \gamma$)) scattering and precision optical tests \cite{NiauAkmansoy:2018ilv,Fouche:2016qqj}. Finally, the broader landscape of singularity resolution includes higher-derivative/gravity-sector mechanisms and alternative nonsingular compact objects (e.g., regular-center constructions and black-bounce-type solutions), offering complementary pathways beyond pure NLED regularization \cite{Cano:2020ezi,Maeda:2021jdc,Junior:2025sjr}.

A central limitation of much of the existing lensing/birefringence literature is the reliance on the scattering limit, where both the source and the observer are placed at spatial infinity. While convenient, that approximation can be poorly matched to compact-object environments in which the dominant contribution to birefringence is accumulated near the emission region, where field gradients are largest. In particular, for magnetars the relevant rays originate deep in the strong-field zone and then sample only an \emph{outgoing} segment of the would-be scattering trajectory, so the observable birefringence is not captured by simply taking ``half'' of an asymptotic result. This motivates a formulation in which finite source and receiver locations are treated exactly at the level of the geometry.

In this work we develop such a finite-distance framework for vacuum birefringence in strong gravitational and electromagnetic fields. Our starting point is the Gauss-Bonnet method for light deflection, which expresses the bending angle as a global curvature integral over an appropriate domain in the optical manifold \cite{Gibbons:2008rj}. We generalize this approach to the birefringent setting by constructing the two polarization-dependent optical geometries and by adopting the finite-distance prescription for geometric angles at the source and observer developed in the gravitational-lensing context \cite{Ishihara:2016vdc,Ishihara:2016sfv,Rindler:2007zz}. This yields a clean, coordinate-invariant representation of the deflection for each polarization mode that remains valid when the endpoints are at finite radius.

A key conceptual and practical ingredient of our analysis is a weak-coupling perturbative expansion in the NLED parameter, which allows us to disentangle the purely gravitational (background) deflection from the genuinely polarization-dependent contribution induced by vacuum nonlinearity. In doing so we identify two distinct mechanisms contributing to the observable birefringence: (i) a \emph{refractive-curvature} term driven by the difference of Gaussian curvatures of the two optical manifolds, and (ii) a \emph{path-induced} term originating from the fact that the two polarization modes trace slightly different spatial trajectories and therefore enclose different curvature flux even in the same background gravitational field. This decomposition makes transparent why finite-distance effects can be sizable: the Gauss-Bonnet domain is effectively truncated when the ray begins at finite radius, and the truncation competes with the steep radial falloff of the NLED-induced curvature.

We apply the general formalism to two benchmark theories: the Euler-Heisenberg effective action of QED and Born-Infeld (BI) electrodynamics \cite{Heisenberg:1936nmg,Schwinger:1951nm,Born:1933pep,Born:1934gh}, used here as a \emph{non-birefringent} consistency check. We derive closed expressions and controlled series expansions for the differential bending angle that explicitly retain the finite positions of the source and observer. In particular, with our definition of vacuum birefringence as the differential deflection
$\Delta\alpha\equiv \alpha_{+}-\alpha_{-}$,
Born-Infeld theory yields identically $\Delta\alpha_{\rm BI}=0$ (the two polarization modes share a single characteristic cone) \cite{Boillat:1970gw,Obukhov:2002xa}.
Finally, specializing to magnetar-motivated configurations, we quantify the observational impact of finite-distance truncation by introducing a geometric suppression factor for surface (or near-surface) emission. We show that the birefringence predicted by asymptotic scattering models can be substantially reduced for realistic emission geometries, reaching order-unity corrections (up to $\sim 50\%$ in representative limb-emission configurations), and we emphasize that incorporating these effects is essential for faithful inference of magnetic-field strengths and birefringent NLED/QED parameters from X-ray polarimetry data \cite{Weisskopf_2016,eXTP:2018anb}.

The paper is organized as follows. In Sec. \ref{sec2} we review the effective-metric construction for polarized photon propagation in NLED backgrounds and define the associated optical manifolds. In Sec. \ref{sec3} we formulate the finite-distance Gauss-Bonnet expression for the deflection angle for each polarization mode. In Sec. \ref{sec4} we develop the perturbative expansion and derive the general birefringence formula, separating refractive-curvature and path-induced contributions. In Sec. \ref{sec5} we apply the framework to Euler-Heisenberg QED and Born-Infeld theory and discuss magnetar-relevant suppression factors. We conclude in Sec. \ref{sec6} with implications for strong-field tests of QED and directions for extensions beyond spherical symmetry.

\section{The geometric foundation of photon propagation} \label{sec2}
In this section, we establish the geometric framework necessary to evaluate the finite-distance birefringence effects induced by NLED. Our primary objective is to derive the polarization-dependent effective optical metrics, denoted as $\gamma_{ij}^{\pm}$, by analyzing the interaction between the background spacetime geometry and the non-linear electromagnetic Lagrangian, which provides the rigorous basis for applying the Gauss-Bonnet theorem in subsequent sections.

\subsection{The background and the field} \label{sec2.1}
We begin by defining the background geometry in which the electromagnetic field propagates. We consider a static, spherically symmetric (SSS) spacetime, which describes the gravitational field of a compact object such as a black hole or a magnetar. Following the standard formalism for SSS solutions, the line element is given by:
\begin{equation}
ds^{2} = g_{\mu\nu}dx^{\mu}dx^{\nu} = -A(r)dt^{2} + B(r)dr^{2} + r^{2}d\Omega^{2}, \label{1.1}
\end{equation}
where the metric functions $A(r)$ and $B(r)$ depend solely on the radial coordinate $r$. The angular element is defined as $d\Omega^{2} \equiv d\theta^{2} + \sin^{2}\theta d\phi^{2}$. Equation \eqref{1.1} serves as the background gravitational field and is assumed to be a solution to the Einstein field equations coupled to an electromagnetic source.

We introduce the electromagnetic field tensor $F_{\mu\nu}$, defined as the antisymmetrized derivative of the gauge potential, $F_{\mu\nu} = \partial_{\mu}A_{\nu} - \partial_{\nu}A_{\mu}$. To preserve the static and spherical symmetry of the background spacetime, the electromagnetic field tensor must share the isometries of the metric $g_{\mu\nu}$. Consequently, the independent non-vanishing field components are restricted to the radial electric field $E(r) = F_{tr}$ and the radial magnetic field strength $\mathcal{B}(r)\equiv F_{\theta\phi}/\sin\theta$.

We formally encapsulate the geometric and field-theoretic inputs of our model in the following definitions. 
We adopt metric signature $(-,+,+,+)$ and define the dual field strength by
$\tilde F^{\mu\nu}\equiv \frac{1}{2}\epsilon^{\mu\nu\rho\sigma}F_{\rho\sigma}$, with $\epsilon^{tr\theta\phi}=+1/\sqrt{-g}$.
The two electromagnetic invariants are
$F\equiv \frac{1}{4}F_{\mu\nu}F^{\mu\nu}$ and $G\equiv \frac{1}{4}F_{\mu\nu}\tilde F^{\mu\nu}$.
For a static, spherically symmetric background with radial fields $E(r)=F_{tr}$ and
$\mathcal{B}(r)\equiv F_{\theta\phi}/\sin\theta$, these reduce to:
\begin{align}
g_{\mu\nu}&=\mathrm{diag}\!\left(-A(r),\,B(r),\,r^{2},\,r^{2}\sin^{2}\theta\right),\nonumber \\
F&\equiv \frac{1}{4}F_{\mu\nu}F^{\mu\nu}
=\frac{1}{2}\left(\frac{\mathcal{B}(r)^{2}}{r^{4}}-\frac{E(r)^{2}}{A(r)B(r)}\right),\nonumber \\
G&\equiv \frac{1}{4}F_{\mu\nu}\tilde F^{\mu\nu}
=\frac{E(r)\,\mathcal{B}(r)}{r^{2}\sqrt{A(r)B(r)}}.
\label{II.2}
\end{align}
It is important to emphasize that the assumption of a spherically symmetric radial magnetic field ($\mathcal{B}(r) \neq 0$) implies the existence of a magnetic monopole charge $q_m$ \cite{Bronnikov:2000vy}. While astrophysical magnetars possess dipole fields that break spherical symmetry, we employ this monopole configuration as a monopole approximation. It allows us to analytically model the effects of extreme magnetic field intensities on photon propagation within the spherically symmetric framework of the Gauss-Bonnet theorem, serving as a foundational step toward more complex geometries.

\subsection{The NLED Lagrangian and effective geodesics} \label{sec2.2}
Having established the background geometry and the electromagnetic invariants, we now turn to the dynamics of the photon field governed by NLED. We postulate a generic action for the electromagnetic sector, defined by the Lagrangian density $\mathcal{L}(F, G)$:
\begin{equation}
S_\text{EM} = \int d^{4}x \sqrt{-g} \mathcal{L}(F, G). \label{1.4}
\end{equation}
The equations of motion are derived by varying the action with respect to the gauge potential $A_{\mu}$. Defining the constitutive tensor $P^{\mu\nu} \equiv \frac{\partial \mathcal{L}}{\partial F_{\mu\nu}}$, the generalized Maxwell equations take the form:
\begin{equation}
\nabla_{\mu} \left( \mathcal{L}_{F} F^{\mu\nu} + \mathcal{L}_{G} \tilde{F}^{\mu\nu} \right) = 0, \label{1.5}
\end{equation}
where $\mathcal{L}_{F} \equiv \partial \mathcal{L} / \partial F$ and $\mathcal{L}_{G} \equiv \partial \mathcal{L} / \partial G$. Unlike in linear Maxwell theory (where $L=-F=-\tfrac14 F_{\mu\nu}F^{\mu\nu}$), the coefficients $\mathcal{L}_{F}$ and $\mathcal{L}_{G}$ here are functions of the field strength, inducing self-interaction and non-linear propagation effects.

To study photon propagation, we employ the geometric optics approximation. We introduce a perturbation $f_{\mu\nu}$ representing the propagating photon field on top of the background field $F_{\mu\nu}$, such that the total field is $F_{\mu\nu} + f_{\mu\nu}$. Under the eikonal approximation, the perturbation propagates as a rapidly oscillating wave $f_{\mu\nu} = \bar{f}_{\mu\nu} e^{i \theta}$, where $k_{\mu} = \partial_{\mu} \theta$ is the wave vector.

Linearizing the field equations \eqref{1.5} with respect to the perturbation yields a wave equation for the photon. The non-linearity of the Lagrangian results in the background field acting as an effective medium. The principal part of this linearized equation leads to the Fresnel equation for the wave vector $k_{\mu}$ \cite{Boillat:1970gw}. Critically, the non-trivial dependence of $L$ on $F$ and $G$ generically yields a quartic Fresnel equation that factorizes into the product of two quadratic dispersion relations, corresponding to two polarization modes (denoted by $\pm$). In exceptional models (notably Born-Infeld electrodynamics) the two quadratics coincide so the Fresnel polynomial is a perfect square; in that case there is \emph{no} vacuum birefringence (a single effective light cone shared by both polarizations) \cite{Boillat:1970gw,Obukhov:2002xa,Guzman-Herrera:2024fkg}. The factorization implies that photons do not traverse the null geodesics of the background metric $g_{\mu\nu}$ (Eq. \eqref{1.1}). Instead, they follow the null geodesics of polarization-dependent effective metrics, denoted $g_{\mu\nu}^{\text{eff}, \pm}$.

To explicitly derive these metrics, we utilize the formalism established by Novello \textit{et al.} and De Lorenci \textit{et al.} \cite{Novello:1999pg,DeLorenci:2000yh}. For a background geometry defined by Eq. \eqref{II.2} with radial fields, the cross-terms in the Fresnel equation vanish, and the effective metric tensor diagonalizes. The contravariant components for the two polarization modes are given by:
\begin{equation}
g_{\text{eff}, \pm}^{\mu\nu} = \mathcal{L}_{F} g^{\mu\nu} - 4 \Lambda_{\pm} F^{\mu}_{\ \lambda}F^{\nu\lambda},\,
\label{II.5}
\end{equation}
where $\Lambda_{\pm}$ are the eigenvalues of the polarization tensor, which encapsulate the non-linear self-interaction of the field. These eigenvalues are functions of the field invariants and the Lagrangian derivatives:
\begin{equation}
\Lambda_{\pm}
=\frac{1}{2}\left[
L_{FF}+\frac{1}{2}\eta^{2}L_{GG}
\pm
\sqrt{\left(L_{FF}-\frac{1}{2}\eta^{2}L_{GG}\right)^{2}+\eta^{2}L_{FG}^{2}} \right],
\label{II.6}
\end{equation}
where $L_{FF}\equiv \partial^{2}L/\partial F^{2}$, $L_{GG}\equiv \partial^{2}L/\partial G^{2}$, and
$L_{FG}\equiv \partial^{2}L/(\partial F\,\partial G)$ are evaluated on the background, and where we introduced the
dimensionless scalar
\begin{equation}
\eta^{2}\equiv 1+\left(\frac{G}{F}\right)^{2}=\frac{F^{2}+G^{2}}{F^{2}}.
\end{equation}
In particular, for purely magnetic or purely electric backgrounds one has $G=0$ and hence $\eta=1$.
In the Maxwell limit $L=-F$, one has $L_F=-1$ and $L_{FF}=L_{GG}=L_{FG}=0$, implying $\Lambda_{\pm}=0$ and thus
Eq. \eqref{II.5} reduces (up to an overall conformal factor) to $g^{\mu\nu}_{\mathrm{eff},\pm}\propto g^{\mu\nu}$, as expected. In the specific case of purely magnetic or purely electric backgrounds often used in SSS models, these expressions simplify, identifying one mode that propagates along the background light cone (the ordinary ray in linear theory limits) and one that deviates (the extraordinary ray).

\subsection{Construction of the dual optical manifolds} \label{sec2.3}
The effective metric tensor $g_{\mu\nu}^{\text{eff}, \pm}$, defined by the null propagation condition in Eq. \eqref{II.5}, inherits the static and spherical symmetries of the background metric $g_{\mu\nu}$ and the field tensor $F_{\mu\nu}$. Consequently, the effective line element for photon propagation can be diagonalized and written in a general SSS form. We introduce the effective metric functions $A_{\pm}(r)$, $B_{\pm}(r)$, and $H_{\pm}(r)$ to describe the geometry seen by the two polarization modes:
\begin{equation}
ds_{\text{eff}, \pm}^{2} = -A_{\pm}(r)dt^{2} + B_{\pm}(r)dr^{2} + H_{\pm}(r)r^{2}d\Omega^{2}, \label{1.8}
\end{equation}
where $H_{\pm}(r)$ accounts for the fact that the angular part of the effective metric may differ from the background $r^2$ due to the non-linear interaction terms $\mathcal{L}_{FF}$ and $\mathcal{L}_{GG}$ identified in Eq. \eqref{II.5}. Note that in the limit of linear electrodynamics, $A_{\pm} \to A$, $B_{\pm} \to B$, and $H_{\pm} \to 1$. Note that since the effective metric tensor derived from the Fresnel equation (Eq. \eqref{II.5}) is contravariant, the functions appearing in the line element (Eq. \eqref{1.8}) are the inverses of the diagonal components: $A_{\pm}(r) = -(g^{tt}_{\text{eff}, \pm})^{-1}$ and $B_{\pm}(r) = (g^{rr}_{\text{eff}, \pm})^{-1}$. The function $H_{\pm}(r)$ encapsulates the deformation of the angular sector relative to the radial sector induced by the non-linearities. Explicitly, it is defined by the ratio of the effective metric components normalized by the background:
\begin{equation}
H_{\pm}(r) = \frac{g^{\theta\theta}_{\text{eff}, \pm} / g^{\theta\theta}}{g^{rr}_{\text{eff}, \pm} / g^{rr}},
\end{equation}
and this factor is unity in linear electrodynamics (Maxwell theory) but deviates in the strong-field regime, driving the birefringence.

To study the spatial trajectory of light rays, we restrict our analysis to the equatorial plane ($\theta = \pi/2$) without loss of generality. By setting the effective line element $ds_{\text{eff}, \pm}^{2} = 0$, we can solve for the coordinate time interval $dt$ in terms of the spatial coordinates, which allows the projection of the four-dimensional null geodesics onto three-dimensional spatial curves. The resulting quadratic form defines the optical metric $\gamma_{ij}^{\pm}$ for each polarization mode \cite{Birula_1986}:
\begin{equation}
dt^{2} = \gamma_{ij}^{\pm} dx^{i} dx^{j} = \frac{B_{\pm}(r)}{A_{\pm}(r)} dr^{2} + \frac{H_{\pm}(r)r^{2}}{A_{\pm}(r)} d\phi^{2}. \label{1.9}
\end{equation}
Equation \eqref{1.9} defines two distinct three-dimensional Riemannian manifolds, which we denote as $M^{opt}_{+}$ and $M^{opt}_{-}$. These manifolds are the optical spaces in which the photons travel as spatial geodesics. The non-vanishing components of the optical metric tensors $\gamma_{ij}^{\pm}$ on the equatorial plane are explicitly identified as:
\begin{equation}
\gamma_{rr}^{\pm}(r) = \frac{B_{\pm}(r)}{A_{\pm}(r)}, \quad \gamma_{\phi\phi}^{\pm}(r) = \frac{H_{\pm}(r)r^{2}}{A_{\pm}(r)}.
\label{1.10}
\end{equation}
The birefringence of the vacuum is geometrically manifest in the fact that $M^{opt}_{+} \neq M^{opt}_{-}$. A single spacetime event emits photons that traverse distinct optical manifolds depending on their polarization, accumulating different geometric phases and deflection angles. These optical metrics $\gamma_{ij}^{\pm}$ serve as the fundamental geometric objects for the application of the Gauss-Bonnet theorem in the following section.

\section{The finite-distance Gauss-Bonnet formalism} \label{sec3}
In this section, we formulate the deflection angle $\alpha_{\pm}$ for a finite-distance observer using the topology of the optical manifolds $M^{opt}_{\pm}$ derived in Section \ref{sec2}. Our approach generalizes the work of Gibbons and Werner \cite{Gibbons:2008rj,Ono:2019hkw,Crisnejo:2018ppm}, who established that the asymptotic deflection angle can be computed as a global topological invariant, which is specifically the surface integral of the Gaussian curvature over the optical manifold. We extend this correspondence to the finite-distance regime by rigorously defining the geometric angles at the observer and source locations within the curved optical space, allowing us to quantify the birefringence $\Delta \alpha$ as a differential geometric effect arising from the distinct curvatures $K^{+}$ and $K^{-}$ of the polarization-dependent manifolds.

\subsection{Gaussian curvature of the optical manifolds} \label{sec3.1}
To apply the Gauss-Bonnet theorem, we must first determine the intrinsic curvature of the optical submanifolds defined by the metric in Eq. \eqref{1.10}. The optical metric for the equatorial plane is given by the 2-dimensional line element \cite{Ono:2017pie}:
\begin{equation}
dl_{\pm}^{2} = \gamma_{rr}^{\pm}(r)dr^{2} + \gamma_{\phi\phi}^{\pm}(r)d\phi^{2}, \label{2.1}
\end{equation}
where the metric components are explicitly defined in terms of the effective metric functions as $\gamma_{rr}^{\pm} = B_{\pm}/A_{\pm}$ and $\gamma_{\phi\phi}^{\pm} = r^2 H_{\pm}/A_{\pm}$.

Since the optical metric is diagonal, the non-vanishing Christoffel symbols of the second kind, $\Gamma^{k}_{ij} = \frac{1}{2}\gamma^{kl}(\partial_{i}\gamma_{jl} + \partial_{j}\gamma_{il} - \partial_{l}\gamma_{ij})$, are straightforward to compute. The relevant radial components required for the curvature calculation are:
\begin{equation}
\Gamma^{r}_{rr} = \frac{1}{2\gamma_{rr}^{\pm}}\frac{d\gamma_{rr}^{\pm}}{dr}, \quad \Gamma^{r}_{\phi\phi} = -\frac{1}{2\gamma_{rr}^{\pm}}\frac{d\gamma_{\phi\phi}^{\pm}}{dr}, \quad \Gamma^{\phi}_{r\phi} = \frac{1}{2\gamma_{\phi\phi}^{\pm}}\frac{d\gamma_{\phi\phi}^{\pm}}{dr}. \label{2.2}
\end{equation}
The Gaussian curvature $K$ for a 2-dimensional surface is an intrinsic scalar related to the Riemann curvature tensor by $K = R_{r\phi r\phi} / \det(\gamma)$. For our orthogonal coordinates, this reduces to a standard differential form involving the metric components:
\begin{equation}
K^{\pm}(r) = -\frac{1}{\sqrt{\gamma_{rr}^{\pm}\gamma_{\phi\phi}^{\pm}}} \frac{d}{dr} \left( \frac{1}{\sqrt{\gamma_{rr}^{\pm}}} \frac{d\sqrt{\gamma_{\phi\phi}^{\pm}}}{dr} \right). \label{2.3}
\end{equation}
Substituting the expressions for $\gamma_{rr}^{\pm}$ and $\gamma_{\phi\phi}^{\pm}$ from Eq. \eqref{1.10} into Eq. \eqref{2.3}, we derive the explicit form of the Gaussian curvature for each polarization mode. Rather than expanding the derivatives fully, it is physically more insightful to present the curvature in a compact factorized form, which reveals how the non-linear effective metric functions modify the standard geometric curvature:
\begin{equation}
K^{\pm}(r) = -\frac{A_{\pm}(r)}{r \sqrt{B_{\pm}(r)H_{\pm}(r)}} \frac{d}{dr} \left( \sqrt{\frac{A_{\pm}(r)}{B_{\pm}(r)}} \frac{d}{dr}\left( r \sqrt{\frac{H_{\pm}(r)}{A_{\pm}(r)}} \right). \right)
\label{2.4}
\end{equation}
Equation \eqref{2.4} explicitly demonstrates that the birefringence arises from the non-trivial radial gradients of the ratio $H_{\pm}/A_{\pm}$ and the conformal factor $A_{\pm}/B_{\pm}$. In the weak-field limit where $A_{\pm} \approx 1 - 2M/r$ and $H_{\pm} \approx 1$, this expression reduces to the standard Schwarzschild optical curvature $K \approx -2M/r^3$. However, in the strong-field regime near the photon sphere, the term $H_{\pm}(r)$ (containing $\mathcal{L}_{FF}$ and $\mathcal{L}_{GG}$) introduces significant deviations.

It is crucial to analyze the behavior of $K^{\pm}$ near the photon sphere radius $r_{ph}^{\pm}$. The photon sphere is defined by the singularity of the effective potential, or equivalently, where the optical metric components diverge or the geodesic equation allows for circular orbits \cite{Li:2020wvn}. If $K^{\pm}$ possesses a singularity at $r_{ph}^{\pm}$, the Gauss-Bonnet integral must be handled with regularization techniques. For standard NLED models like Euler-Heisenberg, the curvature remains regular outside the horizon, but for models with Born-Infeld type structures, divergences may appear that require restricting the integration domain to $r > r_{ph}^{\pm}$. The behavior of the divergence fundamentally distinguishes the NLED optical manifold from the vacuum Schwarzschild case. In the remainder of this work we restrict to the weak-deflection regime ($b\gg M$ and $r_0\simeq b$ well outside
$r_{\rm ph}^\pm$), so the near-photon-sphere regularization issues mentioned here do not enter our analytic evaluation;
extending the construction to strong deflection is left for future work.

\subsection{The finite-distance geometric angles} \label{sec3.3}
With the curvature of the optical manifolds established, we now define the geometric configuration of the lensing system at finite distances. We consider a light source located at a radial coordinate $r_\text{S}$ and a receiver (observer) located at $r_\text{R}$. Even when the background and effective geometries are asymptotically flat in the usual sense, realistic lensing/polarimetry
setups place the source $S$ and receiver $R$ at finite radii $r_S$ and $r_R$, so the endpoint angles $\Psi_{S,R}$ must be
evaluated at finite distance rather than by assuming $r_S,r_R\to\infty$.

We define the local angle $\Psi^{\pm}$ as the angle between the tangent vector of the light ray, $K^{i} = dx^{i}/dt$, and the radial unit vector, $e_{rad}^{i}$, measured within the optical manifold $M^{opt}_{\pm}$. The local angle represents the deviation of the photon's trajectory from the radial direction as measured by a local static observer. In terms of the optical metric components from Eq. \eqref{1.10}, the cosine of this angle is given by the inner product $\cos \Psi^{\pm} = \gamma_{rr}^{\pm} (e_{rad}^{r} K^{r})$.

However, for practical calculations of the deflection angle, the sine of this angle is more convenient as it relates directly to the impact parameter $b$. In a static, spherically symmetric spacetime, the impact parameter is a constant of motion defined by the ratio of the specific angular momentum to the specific energy, $b \equiv L/E$. In terms of the metric functions and coordinate velocities, this is expressed as:
\begin{equation}
b = \frac{\gamma_{\phi\phi}^{\pm}}{\sqrt{\gamma_{ii}^{\pm} (dx^i/dt)(dx^i/dt)}} \frac{d\phi}{dt} \Big|_\text{null} = \frac{H_{\pm}(r) r^2}{A_{\pm}(r)} \frac{d\phi}{dt}, \label{2.5}
\end{equation}
where we have utilized the effective metric decomposition from Eq. \eqref{1.8}.

By projecting the tangent vector onto the orthonormal angular basis vector $e_{ang}^{i} = (0, 1/\sqrt{\gamma_{\phi\phi}^{\pm}})$, we derive the expression for $\sin \Psi^{\pm}$. Substituting the optical metric component $\gamma_{\phi\phi}^{\pm} = H_{\pm}r^2/A_{\pm}$ into the projection yields the generalized formula for the finite-distance angles:
\begin{equation}
\sin \Psi_\text{R,S}^{\pm} = \frac{b}{r_\text{R,S}} \sqrt{\frac{A_{\pm}(r_\text{R,S})}{H_{\pm}(r_\text{R,S})}}.
\label{2.6}
\end{equation}

Note that Equation \eqref{2.6} provides the magnitude of the sine of the angle. In the application of the Gauss-Bonnet theorem to the quadrilateral domain \cite{Ishihara:2016vdc,Ishihara:2016sfv}, $\Psi$ is defined as the angle between the outgoing radial unit vector and the photon tangent vector. For the receiver, this angle $\Psi_\text{R}$ is acute. However, for the source, the photon is directed inward toward the lens, making the angle obtuse. Therefore, in the sum $\Psi_\text{R} - \Psi_\text{S}$, the source angle is treated as $\Psi_\text{S} = \pi - \arcsin(\dots)$, consistent with the exterior angle formulation

Equation \eqref{2.6} highlights a critical consequence of the non-linear interaction: the geometric angles $\Psi_{R}^{\pm}$ and $\Psi_{S}^{\pm}$ are explicitly polarization-dependent. Unlike in standard General Relativity where the angle depends only on the gravitational potential $A(r)$, the NLED modification introduces the factor $H_{\pm}(r)$ in the denominator. Consequently, a receiver at a fixed coordinate $r_\text{R}$ observing a source at $r_\text{S}$ will measure two distinct image positions for the two polarization modes, even before calculating the integrated deflection along the path. Such a local birefringence at the endpoints contributes to the total observable splitting alongside the integrated curvature effect.

\subsection{The birefringence integral} \label{sec3.4}
Having defined the Gaussian curvature $K^{\pm}$ and the boundary angles $\Psi_\text{R,S}^{\pm}$, we now assemble these components using the Gauss-Bonnet theorem to formulate the total deflection angle. We consider the quadrilateral domain ${}^{\infty}_{R}\Box^{\infty}_{S}$ embedded in the optical manifold $M^{opt}_{\pm}$, which is bounded by the geodesic segment connecting the source $S$ and receiver $R$, two radial geodesics extending from $S$ and $R$ to infinity, and a circular arc at infinity.

The Gauss-Bonnet theorem relates the total geodesic curvature of the boundary and the surface integral of the Gaussian curvature to the Euler characteristic of the domain. For our quadrilateral configuration in the simply connected optical manifold, the theorem yields the relation:
\begin{equation}
\iint_{{}^{\infty}_{R}\Box^{\infty}_{S}} K^{\pm} dS^{\pm} + \int_{\partial ({}^{\infty}_{R}\Box^{\infty}_{S})} \kappa_g d\ell + \sum_{i} \theta_i = 2\pi, \label{2.7}
\end{equation}
where $\kappa_g$ is the geodesic curvature of the boundary and $\theta_i$ are the exterior angles at the vertices.

Following the geometric construction by Ishihara \textit{et al.} \cite{Ishihara:2016vdc}, the sum of the jump angles and the boundary integrals can be related to the geometric deflection angle $\alpha_{\pm}$. Specifically, the definition of the deflection angle $\alpha$ given in Eq. (1.17) of the reference text \cite{Rindler:2007zz,Ishihara:2016vdc,Takahashi:2023eli} (and generalized here for polarization modes) is:
\begin{equation}
\alpha_{\pm} \equiv \Psi_{R}^{\pm} - \Psi_{S}^{\pm} + \phi_{RS}^{\pm}, \label{2.8}
\end{equation}
where $\phi_{RS}^{\pm}$ is the coordinate separation angle between source and receiver.

By applying the theorem to the quadrilateral domain ${}^{\infty}_{R}\Box^{\infty}_{S}$, the deflection angle is expressed as the negative of the surface integral of the Gaussian curvature over this oriented area, providing the master equation for the finite-distance deflection of each polarization mode:
\begin{equation}
\alpha_{\pm} = -\iint_{{}^{\infty}_{R}\Box^{\infty}_{S}} K^{\pm} dS^{\pm}.\label{2.9}
\end{equation}

The differential deflection angle, or finite-distance birefringence, is then the difference between the angles for the two modes:
\begin{equation}
\Delta \alpha = \alpha_{+} - \alpha_{-} = -\left( \iint_{D^+} K^{+} dS^{+} - \iint_{D^-} K^{-} dS^{-} \right), \label{2.10}
\end{equation}
where $D^{\pm}$ denotes the integration domain ${}^{\infty}_{R}\Box^{\infty}_{S}$ on the respective optical manifold.

Equation \eqref{2.10} presents a significant calculational challenge: the integration domains $D^+$ and $D^-$ are physically distinct because the photon paths $\gamma^+$ and $\gamma^-$ differ due to the effective metric splitting. To compute $\Delta \alpha$, one cannot simply subtract the integrands. Instead, we must account for both the difference in the curvature scalars ($K^+ - K^-$) and the difference in the integration areas ($\delta S$). This requires a perturbative approach to map both domains onto a common reference manifold (the background geometry), which we will develop in the next section.

\section{Perturbative expansion and differential bending} \label{sec4}
In the previous section, we established that the finite-distance birefringence $\Delta \alpha$ is given by the difference of two area integrals over distinct domains in the optical manifolds $M^{opt}_{\pm}$. Direct integration of Eq. \eqref{2.10} is analytically intractable due to the complex dependencies of the effective metric functions $A_{\pm}$ and $H_{\pm}$ on the field invariants. To overcome this, we employ a perturbative approach. We assume the NLED coupling is weak, which is characterized by a small parameter $\beta$ (e.g., proportional to the fine-structure constant $\alpha_\text{EM}^2$ in QED), allowing us to linearize the geometric quantities around the background General Relativistic (GR) solution. The procedure decouples the gravitational deflection from the NLED-induced birefringence, enabling us to isolate the specific finite-distance corrections arising from the photon's polarization.

\subsection{The weak-field expansion} \label{sec4.1}
We begin by expanding the effective metric functions defined in Eq. \eqref{1.8} with respect to the NLED coupling parameter $\beta$. In the limit $\beta \to 0$, we recover the background metric functions $A(r)$ and $B(r)$, and the angular deformation factor $H(r)$ approaches unity. We write the expansions as:
\begin{equation}
A_{\pm}(r)=A(r)\Big(1+\beta\,a_{\pm}(r)\Big),\qquad
B_{\pm}(r)=B(r)\Big(1+\beta\,b_{\pm}(r)\Big),\qquad
H_{\pm}(r)=1+\beta\,h_{\pm}(r),
\label{IV.1}
\end{equation}
where $a_{\pm}$, $b_{\pm}$, and $h_{\pm}$ are dimensionless $\mathcal{O}(1)$ perturbation functions determined by the
specific NLED Lagrangian (Eq. \eqref{II.5}). (The symbol $b_{\pm}$ should not be confused with the impact parameter $b$.)

Substituting these expansions into the general expression for the Gaussian curvature (Eq. \eqref{2.4}), we linearize $K^{\pm}(r)$. The curvature decomposes into the background gravitational curvature $K_\text{GR}(r)$ and a polarization-dependent correction:
\begin{equation}
K^{\pm}(r) = K_\text{GR}(r) + \beta K^{(1)}_{\pm}(r) + O(\beta^2). \label{3.2}
\end{equation}
Here, $K_\text{GR}$ is the Gaussian curvature of the optical manifold for the background metric (e.g., $K \approx -2M/r^3$ for Schwarzschild), and $K^{(1)}_{\pm}$ encapsulates the NLED contribution involving radial derivatives of the perturbation functions
$a_{\pm}$, $b_{\pm}$, and $h_{\pm}$.

To compute the integrals in Eq. \eqref{2.10}, we must also define the integration domain boundaries. The upper boundary of the quadrilateral ${}^{\infty}_{R}\Box^{\infty}_{S}$ is defined by the photon's spatial trajectory $r(\phi)$. Since the photons follow geodesics of different effective metrics, their trajectories differ. We parameterize the orbit using the inverse radial coordinate $u(\phi) \equiv 1/r(\phi)$. The orbit equation (corresponding to Eq. (11) in Ishihara \textit{et al.}, modified here for the effective metric) is:
\begin{equation}
\left( \frac{du}{d\phi} \right)^2 + \frac{u^2 H_{\pm}}{B_{\pm}} = \frac{1}{b^2} \frac{H_{\pm}}{A_{\pm} B_{\pm}}. \label{3.3}
\end{equation}

We expand the solution $u_{\pm}(\phi)$ around the background null geodesic $u_{0}(\phi)$. For a static spherically symmetric background, the zero-order path is $u_{0}(\phi) = \frac{\sin\phi}{b}$ (assuming the straight-line approximation locally or the exact Schwarzschild solution). The perturbed trajectory is:
\begin{equation}
u_{\pm}(\phi) = u_{0}(\phi) + \beta \delta u_{\pm}(\phi). \label{3.4}
\end{equation}
Substituting the metric expansions (Eq. \eqref{IV.1}) into the orbit equation (Eq. \eqref{3.3}) and keeping terms to first order in $\beta$, we find that the perturbation $\delta u_{\pm}$ satisfies the linearized Binet equation:
\begin{equation}
\frac{d^2 \delta u_{\pm}}{d\phi^2} + \delta u_{\pm} = S_{\pm}(\phi),
\end{equation}
where the source term $S_{\pm}$ depends on the derivatives of the effective metric perturbation functions. Unlike the background equation, which may possess singularities at the turning point if solved via direct integration, this linear ODE describes a driven oscillator. The physical solution, satisfying the boundary condition $\delta u(0) = 0$ (assuming asymptotic alignment at the observer), is constructed using the Green's function $G(\phi, \varphi) = \sin(\phi - \varphi)$:
\begin{equation}
\delta u_{\pm}(\varphi)=\int_{0}^{\varphi}\sin(\varphi-\phi)\,
\Big[u_{0}(\phi)^{2}\big(h_{\pm}'(u_{0})-a_{\pm}'(u_{0})-2\,b_{\pm}(u_{0})\big)\Big]\;d\phi.
\label{IV.6}
\end{equation}
Here a prime denotes differentiation with respect to $u$ evaluated along the background orbit $u=u_{0}(\phi)$. The integral in Eq. \eqref{IV.6} describes the physical deviation of the photon path from the background null geodesic due to the effective NLED force. Crucially, this form remains regular at the turning point of the orbit, allowing for a consistent evaluation of the path-induced birefringence term.

The perturbation $\delta u_{\pm}$ is critical: it signifies that for a fixed coordinate angle $\phi$, the two polarization modes are located at different radial positions $r_+ \neq r_-$. The spatial splitting implies that the integration domains $D^+$ and $D^-$ in Eq. \eqref{2.10} do not overlap perfectly. In the next subsection, we will utilize this trajectory shift to compute the area term of the differential bending angle.

\subsection{The differential bending angle} \label{sec4.2}
With the trajectory perturbation $\delta u_{\pm}$ determined, we proceed to calculate the differential deflection angle $\Delta \alpha$, which constitutes the observable birefringence. As established in Eq. \eqref{2.10}, $\Delta \alpha$ is defined by the difference between the Gauss-Bonnet integrals over the two polarization-dependent domains $D^+$ and $D^-$.

To make the calculation tractable, we linearize the integrals with respect to the coupling parameter $\beta$. The difference can be decomposed into two distinct physical contributions: a bulk curvature contribution arising from the difference in the effective refractive indices (the integrands), and a boundary contribution arising from the spatial separation of the photon paths (the domains).

We express the area element difference as $\Delta(dS) = dS^+ - dS^- \approx \beta \frac{\partial}{\partial \beta}(\sqrt{\gamma^{\pm}}) dr d\phi$. Similarly, the curvature difference is $\Delta K = K^+ - K^- \approx \beta (K^{(1)}_+ - K^{(1)}_-)$.

The domains $D^+$ and $D^-$ share the same asymptotic boundaries at infinity but differ at the inner boundary defined by the photon orbits $r_{\pm}(\phi) = 1/u_{\pm}(\phi)$. The integration with respect to the radial coordinate extends from the photon path to infinity, i.e., $\int_{r_{\pm}(\phi)}^{\infty}$. Using the Leibniz integral rule, the variation of the domain introduces a boundary term proportional to the shift in the lower limit of integration.

Combining these expansions, the differential birefringence angle is derived as:
\begin{equation}
\Delta \alpha = - \left[ \iint_{D_0} \left( K^+ \sqrt{\gamma^+} - K^- \sqrt{\gamma^-} \right) dr d\phi - \int_{\phi_\text{S}}^{\phi_\text{R}} \left( K_\text{GR}(r) \sqrt{\gamma_\text{GR}(r)} \right)\bigg|_{r=r_0(\phi)} \Delta r(\phi) d\phi \right], \label{3.6}
\end{equation}
where $\Delta r(\phi) \approx -(\delta u_+ - \delta u_-)/u_0^2$ represents the radial separation of the rays. Note that a positive $\delta u$ (larger inverse radius) corresponds to a photon path closer to the lens (smaller $r$).

The second term in Eq. \eqref{3.6} is the area shift correction. Using the perturbation relation $\Delta r \approx - \Delta (\delta u) / u_0^2$ derived from Eq. \eqref{IV.6}, we can express the total finite-distance birefringence explicitly in terms of the NLED perturbations. The formula isolates the NLED effects from the background gravitational lensing:
\begin{equation}
\Delta \alpha = - \underbrace{\int_{\phi_\text{S}}^{\phi_\text{R}} \int_{1/u_0(\phi)}^{\infty} \Delta \left( K \sqrt{\gamma} \right) dr d\phi}_{\text{Refractive Curvature Difference}} - \underbrace{\int_{\phi_\text{S}}^{\phi_\text{R}} \left[ \frac{K_\text{GR}}{u_0^2} \sqrt{\gamma_\text{GR}} \right]_{r=1/u_0} (\delta u_+ - \delta u_-) d\phi}_{\text{Path-Induced Birefringence}}.
\label{IV.8}
\end{equation}
The second term in Eq. \eqref{IV.8} is a \emph{boundary (domain-shift) correction}: it arises when the two Gauss-Bonnet area
integrals over $D_+$ and $D_-$ are mapped onto the same reference domain $D_0$ bounded by the background orbit
$r_0(\phi)=1/u_0(\phi)$. Concretely, for any smooth integrand $\mathcal{I}(r,\phi)$ one has, by the Leibniz rule,
\begin{equation}
\int_{\phi_S}^{\phi_R}\!\!\int_{r_\pm(\phi)}^\infty \mathcal{I}(r,\phi)\,dr\,d\phi
=
\int_{\phi_S}^{\phi_R}\!\!\int_{r_0(\phi)}^\infty \mathcal{I}(r,\phi)\,dr\,d\phi
-\int_{\phi_S}^{\phi_R}\left.\mathcal{I}(r,\phi)\right|_{r=r_0(\phi)}\delta r_\pm(\phi)\,d\phi
+\mathcal{O}(\beta^2),
\tag{IV.8a}
\label{IV.8a}
\end{equation}
with $\delta r_\pm(\phi)\equiv r_\pm(\phi)-r_0(\phi)\simeq-\delta u_\pm(\phi)/u_0(\phi)^2$. Applying
Eq. \eqref{IV.8a} to $\mathcal{I}=K\sqrt{\gamma}$ and subtracting the $\pm$ expressions reproduces Eq. \eqref{IV.8}
identically to $\mathcal{O}(\beta)$. In this sense, the split in Eq. \eqref{IV.8} is a convenient bookkeeping device tied to the
choice of reference curve $u_0$: each term separately changes if $u_0$ is shifted by a common $\mathcal{O}(\beta)$ amount,
but their sum (the observable $\Delta\alpha$) is invariant to the working order.

Physically, the boundary term measures the change in the \emph{background-curvature} flux induced by the
polarization-dependent displacement of the photon orbit. A nonzero contribution therefore requires
$\delta u_+\neq \delta u_-$, i.e.\ genuinely distinct polarization-dependent geodesics. While one can construct special cases
in which the bulk integrand difference $\Delta(K\sqrt{\gamma})$ is suppressed on $D_0$ yet
$\delta u_+-\delta u_-\neq0$, this requires a non-generic cancellation in the curvature/area densities; in typical NLED
models the same mode dependence that produces trajectory splitting also yields $\Delta(K\sqrt{\gamma})\neq 0$.
In particular, for an exceptional non-birefringent theory in which $g^{\mu\nu}_{\rm eff,+}$ and $g^{\mu\nu}_{\rm eff,-}$
coincide (up to a common conformal rescaling) one has $\delta u_+=\delta u_-$ and hence the boundary contribution
vanishes identically; Born-Infeld electrodynamics is of this type \cite{Boillat:1970gw,Obukhov:2002xa}. For finite distances, the limits of the angular integration $\phi_\text{S}$ and $\phi_\text{R}$ (corresponding to $r_\text{S}$ and $r_\text{R}$) are finite, meaning the birefringence accumulation is truncated relative to the asymptotic case. The truncation leads to the finite-distance corrections we aim to isolate in the final subsection.

\subsection{The finite-distance correction series} \label{sec4.3}
The expression derived in Eq. \eqref{IV.8} provides the exact first-order differential birefringence angle. However, to make the result useful for observational tests (e.g., with magnetars), we must separate the result into the standard asymptotic value (observable at infinity) and the finite-distance corrections that arise when the observer or source is near the lens.

We perform a series expansion of the integration limits and the integrands in powers of the inverse radial coordinates $u_\text{R} = 1/r_\text{R}$ and $u_\text{S} = 1/r_\text{S}$. The angular limits $\phi_\text{R}$ and $\phi_\text{S}$ are related to the radial coordinates via the background orbit equation $u_0(\phi) = \sin\phi/b$. In the weak-field limit, this implies $\phi_\text{R} \approx \arcsin(b u_\text{R})$ and $\phi_\text{S} \approx \pi - \arcsin(b u_\text{S})$.

Expanding the integral in Eq. \eqref{IV.8} around these limits, we isolate the terms that depend on $u_\text{R}$ and $u_\text{S}$. The total differential deflection $\Delta \alpha$ can be written as:
\begin{equation}
\Delta \alpha (r_\text{R}, r_\text{S}) = \Delta \alpha_{\infty} - \left( \delta \alpha(r_\text{R}) + \delta \alpha(r_\text{S}) \right), \label{3.8}
\end{equation}
where $\Delta\alpha_\infty$ denotes the \emph{two-sided asymptotic scattering} value obtained in the limit
$r_S,r_R\to\infty$ (so that $\phi_S\to 0$ and $\phi_R\to\pi$ for fixed impact parameter $b$), i.e. by extending the
integration over the full in-and-out trajectory used in the Ishihara-Ono-Asada-Suzuki construction.

To leading order in the coupling $\beta$ and the gravitational potential $M/b$, the finite-distance correction terms are dominated by the truncation of the refractive curvature integral. Explicitly evaluating the integral for a generic power-law falloff of the effective curvature difference $\Delta (K\sqrt{\gamma}) \sim C_n / r^n$, we find:
\begin{equation}
\delta \alpha(r_\text{R,S}) \approx \int_{0}^{\arcsin(b/r_\text{R,S})} \frac{C_n}{(b/\sin\phi)^n} d\phi. \label{3.9}
\end{equation}

For physically relevant \emph{birefringent} models such as vacuum polarization (Euler-Heisenberg QED), the effective curvature \emph{difference} $\Delta(K\sqrt{\gamma})$ decays rapidly with distance. Consequently, the finite-distance correction can be expressed as a series expansion in terms of the ratio of the impact parameter to the observer distance. The general form of the finite-distance birefringence is:
\begin{equation}
\Delta \alpha(r_\text{R}, r_\text{S}) = \Delta \alpha_{\infty} \left[ 1 - c_\text{R} \left( \frac{b}{r_\text{R}} \right)^n - c_\text{S} \left( \frac{b}{r_\text{S}} \right)^n + \dots \right],
\label{IV.11}
\end{equation}
where the exponent $n$ is fixed by the leading large-$r$ scaling of the effective curvature difference
$\Delta(K\sqrt{\gamma})\sim r^{-n}$ (hence $n$ is \emph{explicit} once the far-field decay of the background field and the NLED
model are specified). By contrast, $c_S$ and $c_R$ are dimensionless $\mathcal{O}(1)$ coefficients that encode the normalization
of $\Delta\alpha_\infty$ (two-sided scattering) and the relative weight of the truncated asymptotic tail compared to the
non-asymptotic part of the curvature flux; they are therefore not fixed by power counting alone. In practice, for a given model
and geometry, $c_S$ and $c_R$ are obtained by matching the asymptotic expansion \eqref{IV.11} to the exact first-order result
\eqref{IV.8} (numerically if needed). In Sec. \ref{sec5} we compute $n$ explicitly for the Euler-Heisenberg monopole benchmark and
quote benchmark values for $c_{S,R}$ in the surface-emission configuration.

\subsection{Domain of validity and expected accuracy}
\label{sec:validity}
Our analytic results combine three controlled approximations, and are intended to be used only within their overlap regime: 
(i) For the weak-NLED expansion, the perturbative mapping in Sec. \ref{sec4} assumes that the polarization-dependent metric deformations are small,
\begin{equation}
|\beta\,a_{\pm}(r)|\ll 1,\qquad |\beta\,b_{\pm}(r)|\ll 1,\qquad |\beta\,h_{\pm}(r)|\ll 1,
\label{eq:validity_nled}
\end{equation}
equivalently that the relevant dimensionless combinations built from the Lagrangian derivatives (e.g.\ $\beta F$, $\beta G$)
remain perturbative along the photon path.
For Euler-Heisenberg QED in a purely magnetic background, this corresponds to the standard condition
$\epsilon_{\rm QED}\sim \frac{\alpha_{\rm EM}}{45\pi}\left(\frac{\mathcal{B}}{B_Q}\right)^2 \ll 1$ (with $B_Q=m_e^2/e$),
so that omitted higher-order terms scale as $\mathcal{O}(\epsilon_{\rm QED}^2)$. (ii) In evaluating the finite-distance coefficients we use the background orbit $u_0(\phi)=\sin\phi/b$ and the associated
limit expansions for $\phi_{S,R}$, which corresponds to retaining the leading weak-deflection geometry.
Corrections from using the exact Schwarzschild background orbit enter parametrically at $\mathcal{O}(M/b)$ in the
geometric limits and at higher order in the curvature flux; thus the leading truncation exponents $n$ are robust, while the
order-unity coefficients $c_{S},c_{R}$ inherit $\mathcal{O}(M/b)$ uncertainties in the weak-field treatment. (iii) Throughout we place the electromagnetic configuration on a fixed background geometry (e.g. Schwarzschild) and neglect
electromagnetic backreaction on the spacetime metric. This is appropriate when the electromagnetic stress-energy is small
compared to the curvature scale set by the compact object (e.g.\ for neutron stars, $\mathcal{B}^2/8\pi\ll \rho c^2$), so that the
dominant gravitational lensing is governed by the stellar mass rather than by the field energy.

For representative magnetar surface fields $\mathcal{B}_{\rm surf}\sim 10^{14}--10^{15}\,\mathrm{G}$, one has $\epsilon_{\rm QED}\lesssim 10^{-2}$-$10^{-1}$,
so the leading-order Euler-Heisenberg birefringent corrections retained here are expected to be accurate at the few-percent
level (with higher-order QED corrections becoming important only for substantially larger $B$).
For gravitational compactness $M/R_{\rm NS}\sim 0.2$, the weak-deflection approximation is not extremely small for
near-surface rays, so precise inference of order-unity coefficients should ultimately use the exact background orbit; the
finite-distance \emph{scaling} and geometric truncation mechanism derived in Secs. \ref{sec3}-\ref{sec4} remain the central qualitative
prediction.

\section{Applications to physical metrics} \label{sec5}
Having established the general perturbative framework for finite-distance birefringence in Section \ref{sec4}, we now apply this formalism to specific physical scenarios of interest. The power of the expression derived in Eq. \eqref{IV.8} lies in its universality; it allows us to compute the observable differential deflection for any NLED theory, provided the background metric and the specific form of the Lagrangian coupling are known. We stress that while the decomposition in Eq. \eqref{IV.8} into a bulk term and a boundary (domain-shift) term depends on the
chosen reference orbit used to linearize the domains, the \emph{sum} equals the direct difference of Gauss-Bonnet integrals
and is therefore the invariant observable $\Delta\alpha$ to $\mathcal{O}(\beta)$. In this section, we compute the explicit finite-distance birefringence for the Euler-Heisenberg effective action of QED \cite{Fu:2021akc,Amaro:2022del}, and we then revisit Born-Infeld electrodynamics as a non-birefringent consistency check. 
With our observable defined as $\Delta\alpha\equiv\alpha_{+}-\alpha_{-}$, Born-Infeld theory gives $\Delta\alpha_{\rm BI}=0$ identically \cite{Boillat:1970gw,Obukhov:2002xa}, while any BI modification of the \emph{individual} bending angles is polarization independent and therefore lies outside $\Delta\alpha$.

\subsection{Case Study 1: Quantum electrodynamics (Euler-Heisenberg)} \label{sec5.1}
The first and most fundamental application of our formalism is to the low-energy effective action of QED. Vacuum birefringence is a long-predicted consequence of the interaction of photons with virtual electron-positron pairs in the presence of strong electromagnetic fields.

\subsubsection{The effective Lagrangian and metric perturbations} \label{sec5.1.1}
We consider the standard Euler-Heisenberg Lagrangian density in the weak-field limit. Expanding to the lowest order in the fine-structure constant $\alpha_\text{EM}$ beyond the Maxwell term, the Lagrangian is given by:
\begin{equation}
\mathcal{L}_\text{EH} = -\frac{1}{4}F_{\mu\nu}F^{\mu\nu} + \frac{\beta_\text{QED}}{4} \left[ (F_{\mu\nu}F^{\mu\nu})^2 + \frac{7}{4} (F_{\mu\nu}\tilde{F}^{\mu\nu})^2 \right], \label{V.1}
\end{equation}
where the coupling parameter is $\beta_\text{QED} = \frac{2\alpha_\text{EM}^2}{45 m_e^4}$, with $m_e$ representing the electron mass. Such a form matches the perturbative structure assumed in Eq. \eqref{IV.1}, where $\beta_\text{QED}$ serves as the expansion parameter.

For the background field configuration, we assume a static, spherically symmetric magnetic monopole solution with magnetic charge $Q_m$ in a Schwarzschild spacetime. While magnetic monopoles remain hypothetical, this configuration is standard in theoretical birefringence studies because it preserves spherical symmetry, greatly simplifying the evaluation of the optical metric functions $A_{\pm}$ and $B_{\pm}$. The background electromagnetic field tensor has a single non-vanishing component $F_{\theta\phi} = Q_m \sin\theta$, corresponding to a radial magnetic field $|\mathcal{B}(r)|\simeq Q_m/r^{2}$.

In this configuration one has
\begin{equation}
F_{\mu\nu}F^{\mu\nu}=\frac{2Q_m^{2}}{r^{4}},\qquad
F_{\mu\nu}\tilde F^{\mu\nu}=0,
\end{equation}
i.e.\ in our invariant conventions (Sec. \ref{sec2}) $F=\tfrac14F_{\mu\nu}F^{\mu\nu}=Q_m^{2}/(2r^{4})$ and $G=0$.

For later use it is convenient to rewrite Eq. \eqref{V.1} in terms of $(F,G)$:
\begin{equation}
L_{\rm EH}(F,G)=-F+\beta_{\rm QED}\bigl(4F^{2}+7G^{2}\bigr).
\end{equation}
Hence
\begin{equation}
L_F=-1+8\beta_{\rm QED}F,\qquad
L_{FF}=8\beta_{\rm QED},\qquad
L_{GG}=14\beta_{\rm QED},\qquad
L_{FG}=0,
\end{equation}
evaluated on the background ($G=0$).

For the monopole background, the only non-vanishing components of the contraction
$T^{\mu\nu}\equiv F^{\mu}{}_{\lambda}F^{\nu\lambda}$ are angular. Explicitly, with
$F_{\theta\phi}=Q_m\sin\theta$ one finds
\begin{equation}
T^{tt}=T^{rr}=0,\qquad
T^{\theta\theta}=\frac{Q_m^{2}}{r^{6}},\qquad
T^{\phi\phi}=\frac{Q_m^{2}}{r^{6}\sin^{2}\theta}.
\end{equation}
Substituting into the effective-metric definition (Sec. \ref{sec2}),
\begin{equation}
g^{\mu\nu}_{{\rm eff},\pm}=L_F\,g^{\mu\nu}-4\Lambda_{\pm}\,T^{\mu\nu},
\end{equation}
shows that (up to the common conformal factor $L_F$) the polarization-dependent corrections enter
\emph{precisely} through the angular components. In particular,
\begin{equation}
\frac{g^{rr}_{{\rm eff},\pm}}{g^{rr}}=L_F,\qquad
\frac{g^{\theta\theta}_{{\rm eff},\pm}}{g^{\theta\theta}}=L_F-4\,\Lambda_{\pm}\frac{Q_m^{2}}{r^{4}},
\end{equation}
and therefore, using the definition of $H_\pm$ (Sec. \ref{sec2}),
\begin{equation}
H_\pm(r)=
\frac{g^{\theta\theta}_{{\rm eff}\,\pm}/g^{\theta\theta}}{g^{rr}_{{\rm eff}\,\pm}/g^{rr}}
=
1-\frac{4\Lambda_{\pm}}{L_F}\frac{Q_m^{2}}{r^{4}}
=
1+\beta_{\rm QED}\,h_\pm(r)+\mathcal{O}(\beta_{\rm QED}^{2}),
\label{V.2}
\end{equation}
with the leading angular-sector perturbation
\begin{equation}
h_\pm(r)=4\,\bar\Lambda_\pm\,\frac{Q_m^{2}}{r^{4}},
\qquad
\text{where}\quad
\Lambda_\pm=\beta_{\rm QED}\bar\Lambda_\pm
\ \text{on the monopole background}. \label{V.9}
\end{equation}
For $G=0$ and $L_{FG}=0$, Eq. \eqref{II.6} reduces to $\Lambda_{+}=L_{FF}$ and $\Lambda_{-}=\tfrac12L_{GG}$, so that for EH
$\bar\Lambda_{+}=8$ and $\bar\Lambda_{-}=7$ in the present conventions. (Any polarization-\emph{independent} rescaling of $A_\pm$ and $B_\pm$ induced by $L_F$ is a common conformal factor
and does not affect null geodesics; the birefringent input relevant to $\Delta\alpha$ is the \emph{mode dependence} of $H_\pm$.)

\subsubsection{The finite-distance birefringence profile} \label{sec5.1.2}
To compute the differential bending angle $\Delta \alpha$, we utilize the decomposition into refractive curvature and path-induced contributions derived in Eq. \eqref{IV.8}. The difference in the effective optical Gaussian curvature $\Delta K$ is the dominant source of birefringence. Given the $1/r^{4}$ scaling of the \emph{angular-sector} perturbations $h_\pm$ in Eq. \eqref{V.9}, the leading curvature
difference scales as the second radial derivative of the corresponding metric deformation:
\begin{equation}
\Delta K(r)\approx
\frac{d^{2}}{dr^{2}}
\!\left(
\beta_{\rm QED}\,\Delta\bar\Lambda\,\frac{Q_m^{2}}{r^{4}}
\right)
\sim
\beta_{\rm QED}\frac{Q_m^{2}}{r^{6}},
\label{V.3}
\end{equation}
where $\Delta\bar\Lambda\equiv \bar\Lambda_{+}-\bar\Lambda_{-}$.
Inserting Eq. \eqref{V.3} into the refractive curvature integral of Eq. \eqref{IV.8}, we integrate over the domain bounded by the photon path $r(\phi) \approx b/\sin\phi$. The area element contributes a factor of $\sqrt{\gamma} \sim r$, making the integrand scale as $\Delta (K\sqrt{\gamma}) \sim r^{-5}$.

Performing the integration over the radial coordinate first, as prescribed in Eq. \eqref{IV.11}, we find that the accumulated birefringence falls off rapidly with the impact parameter $b$. Specifically, the asymptotic birefringence scales as $\Delta \alpha_{\infty} \propto \beta_\text{QED} Q_m^2 / b^4$. The steep fall-off ($n=4$) reflects the rapid decay of the \emph{monopole} (spherically symmetric) vacuum-polarization
contribution in the Euler--Heisenberg benchmark; by contrast, the dipole-like scaling discussed later corresponds to the
steeper index $n=6$.

We can now explicitly write the finite-distance correction for QED vacuum birefringence. Substituting the scaling $n=4$ into the general series solution Eq. \eqref{IV.11}, we obtain:
\begin{equation}
\Delta \alpha_\text{QED}(r_\text{R}, r_\text{S}) = \Delta \alpha_{\infty} \left[ 1 - c_\text{R} \left( \frac{b}{r_\text{R}} \right)^4 - c_\text{S} \left( \frac{b}{r_\text{S}} \right)^4 + \mathcal{O}\left( \frac{b^6}{r^6} \right) \right].
\label{V.4}
\end{equation}
The coefficients $c_\text{R,S}$ are positive, order-unity constants arising from the truncation of the angular integral $\int (\sin\phi)^4 d\phi$.

\subsubsection{Physical implications} \label{sec5.1.3}
Equation \eqref{V.4} highlights a critical constraint for observational campaigns. The correction term scales as the fourth power of the ratio $b/r$. Unlike the gravitational deflection of light (which scales as $b/r$ in the finite-distance correction), the QED birefringence is highly localized near the source. Consequently, for an observer at infinity ($r_\text{R} \to \infty$), the finite-distance correction vanishes rapidly. However, if the source is located within the strong-field region (small $r_\text{S}$, e.g., emission from the surface of a magnetar), the truncation term $c_\text{S} (b/r_\text{S})^4$ becomes significant. The implication is that the standard asymptotic formulas over-estimate the expected polarization rotation for surface emission models, necessitating the use of Eq. \eqref{V.4} for precise polarimetric modeling of magnetars.

\subsection{Case Study 2: The Born-Infeld black hole} \label{sec5.2}
While Quantum Electrodynamics provides the standard birefringent vacuum response at low energies, Born-Infeld (BI) electrodynamics offers a classical non-linear extension of Maxwell's theory, originally proposed to regularize the self-energy of point charges \cite{Kim:2021grj}. 
A key structural property of BI theory is that it is \emph{exceptional}: the two polarization modes share the same characteristic cone (no vacuum birefringence) \cite{Boillat:1970gw,Obukhov:2002xa}. 
We therefore use BI here as a consistency check of our finite-distance Gauss-Bonnet framework, emphasizing that our observable is the \emph{differential} deflection $\Delta\alpha\equiv\alpha_{+}-\alpha_{-}$.

\subsubsection{Lagrangian and weak-field limit} \label{sec5.2.1}
The Born-Infeld Lagrangian density is given by
\begin{equation}
L_{\rm BI}=\gamma^{2}\left[1-\sqrt{1+\frac{1}{2\gamma^{2}}F_{\mu\nu}F^{\mu\nu}-\frac{1}{16\gamma^{4}}\left(F_{\mu\nu}\tilde F^{\mu\nu}\right)^{2}}\right],
\label{V.5}
\end{equation}
where $\gamma$ is the Born-Infeld coupling parameter (with dimensions of field strength). In the limit $\gamma\to\infty$ we recover Maxwell's theory. To apply the weak-coupling expansion of Sec. \ref{sec4} we identify $\beta_{\rm BI}\propto \gamma^{-2}$.

Expanding Eq. \eqref{V.5} for weak fields and retaining terms through ${\cal O}(\gamma^{-2})$, we obtain
\begin{equation}
L_{\rm BI}\approx -\frac{1}{4}F_{\mu\nu}F^{\mu\nu}
+\frac{1}{32\gamma^{2}}
\Big[\big(F_{\mu\nu}F^{\mu\nu}\big)^{2}+\big(F_{\mu\nu}\tilde F^{\mu\nu}\big)^{2}\Big]
+{\cal O}\!\left(\gamma^{-4}\right).
\label{V.6}
\end{equation}
Although $F_{\mu\nu}\tilde F^{\mu\nu}=0$ for purely electric or purely magnetic backgrounds, the full two-invariant structure of BI theory is essential for its characteristic (light-cone) geometry \cite{Boillat:1970gw,Obukhov:2002xa}.

\subsubsection{Absence of birefringence} \label{sec5.2.2}
Born-Infeld theory is non-birefringent. That is, the Fresnel polynomial degenerates to a perfect square, so the two polarization modes propagate on the same effective null cone (up to conformal rescaling) \cite{Boillat:1970gw,Obukhov:2002xa}. In our notation this implies
\begin{equation}
g^{\mu\nu}_{{\rm eff},+}=g^{\mu\nu}_{{\rm eff},-},
\qquad
A_{+}=A_{-},\quad B_{+}=B_{-},\quad H_{+}=H_{-},
\label{V.7}
\end{equation}
and therefore the trajectory perturbations coincide, $\delta u_{+}=\delta u_{-}$. Substituting into the master formula Eq. \eqref{IV.8}, both the refractive-curvature term and the path-induced term vanish, yielding
\begin{equation}
\Delta\alpha_{\rm BI}(r_{R},r_{S})\equiv \alpha_{+}-\alpha_{-}=0.
\label{V.8}
\end{equation}
We stress that BI electrodynamics can still modify the \emph{common} bending angle $\alpha_{+}=\alpha_{-}$ relative to Maxwell/GR, but this polarization-independent shift is not part of $\Delta\alpha$.

\subsubsection{Multipole sensitivity of finite-distance suppression} \label{sec5.2.3}
The finite-distance correction series Eq. \eqref{IV.11} is sensitive to the multipole structure of the background field through the large-$r$ decay of $\Delta(K\sqrt{\gamma})$. For a dipole, the background field falls off as $\mathcal{B}\sim r^{-3}$, implying $F^{2}\sim r^{-6}$; correspondingly the leading birefringent curvature difference scales more steeply than in the monopole benchmark. Inserting this into Eq. \eqref{IV.8} yields the parametric scaling
\begin{equation}
\delta\alpha_{\rm dipole}(r)\sim
\int_{0}^{\arcsin(b/r)}(\sin\phi)^{6}\,d\phi
\quad\Longrightarrow\quad n=6,
\label{V.16}
\end{equation}
demonstrating that the truncation terms $c_{S}(b/r_{S})^{n}$ can become even more important when higher multipoles contribute near the surface.

We have seen that the perturbative formalism successfully isolates the distance-dependent reduction of birefringence. For the monopole benchmark (with $n=4$) the signal is suppressed by a $(b/r)^{4}$ factor for nearby sources, while for dipole-like scaling ($n=6$) the convergence to the asymptotic limit is even faster. This steep dependence implies that X-ray polarimetry of sources at finite distances (such as binary partners or surface emission) must account for these truncation factors to avoid underestimating the intrinsic non-linear coupling parameters.

\subsection{The magnetar observer test} \label{sec5.3}
The finite-distance corrections derived above find their most critical application in the study of magnetars, neutron stars with
ultra-strong magnetic fields ($\mathcal{B}\sim 10^{14}--10^{15}\,\mathrm{G}$). We emphasize, however, that the Gauss-Bonnet construction in
Secs. \ref{sec3}-\ref{sec4} assumes an effective optical metric of static, spherically symmetric (SSS) form. A realistic magnetar magnetosphere
is dipole-dominated and therefore axisymmetric \cite{Kim:2024npq,Kim:2024fle}, so an exact dipole calculation lies outside the present SSS framework.
Accordingly, in this subsection we use the dipole only to extract the radial falloff exponent controlling the truncation series and to
illustrate the purely geometric suppression produced by finite-distance emission, particularly for near-limb rays.

\subsubsection{Geometric configuration and suppression factor} \label{sec5.3.1}
We consider a physical scenario where polarized X-rays are emitted from a hot spot on the surface of a magnetar with radius $R_{NS}$ and observed by a distant telescope ($r_\text{R} \to \infty$). In this limit, the term depending on the observer's position in Eq. \eqref{IV.11} vanishes ($c_R(b/r_R)^n\to 0$). The source term, however, remains dominant.

For a photon emitted from the surface at an angle $\psi$ relative to the local radial vector, the impact parameter $b$ is related to the stellar radius by the geometric relation (in the weak-gravity limit):
\begin{equation}
b \approx R_{NS} \sin \psi. \label{4.10}
\end{equation}
Note that this geometric relation assumes the weak-gravity limit ($M \ll R_{NS}$), neglecting the gravitational redshift and light bending at the launch point. In a fully relativistic treatment, the impact parameter would be enhanced by the redshift factor $(1-2M/R_{NS})^{-1/2}$.

For a physical magnetar characterized by a dipole field, the background magnetic field decays as $\mathcal{B}\propto r^{-3}$, leading to an energy density scaling of $r^{-6}$. Consequently, the finite-distance suppression factor is significantly steeper than the monopole approximation ($n=4$) derived in the previous subsection. Adapting the general series expansion (Eq. \eqref{IV.11}) to the dipole case ($n=6$), the observable differential deflection angle becomes:
\begin{equation}
\Delta \alpha_{obs} = \Delta \alpha_{\infty} \left[ 1 - c_\text{S} \left( \frac{R_{NS} \sin \psi}{R_{NS}} \right)^6 \right]. \label{V.18}
\end{equation}
For surface emission, the photon originates near the periastron of the would-be scattering orbit and travels only outward.
If one compares this one-sided trajectory to a \emph{two-sided} scattering orbit through an idealized \emph{symmetric} extension of the
exterior field, the missing inbound segment suggests a truncation coefficient of order unity; in the perfectly symmetric benchmark
one obtains $c_S=1/2$. We therefore adopt $c_S\simeq 1/2$ as an \emph{illustrative} reference value, while stressing that for a realistic
axisymmetric dipole magnetosphere (and for non-equatorial emission) the effective $c_S$ can differ by an $\mathcal{O}(1)$ factor.

With $b\simeq R_{\rm NS}\sin\psi$ this yields the benchmark suppression profile
\begin{equation}
F_{\rm dipole}(\psi)\equiv \frac{\Delta\alpha_{\rm obs}}{\Delta\alpha_\infty}
\simeq 1-\frac{1}{2}\sin^{6}\psi,
\label{V.19}
\end{equation}
which indicates an order-unity reduction for near-limb emission within this symmetric one-sided benchmark.

\begin{figure}
    \centering
    \includegraphics[width=0.6\linewidth]{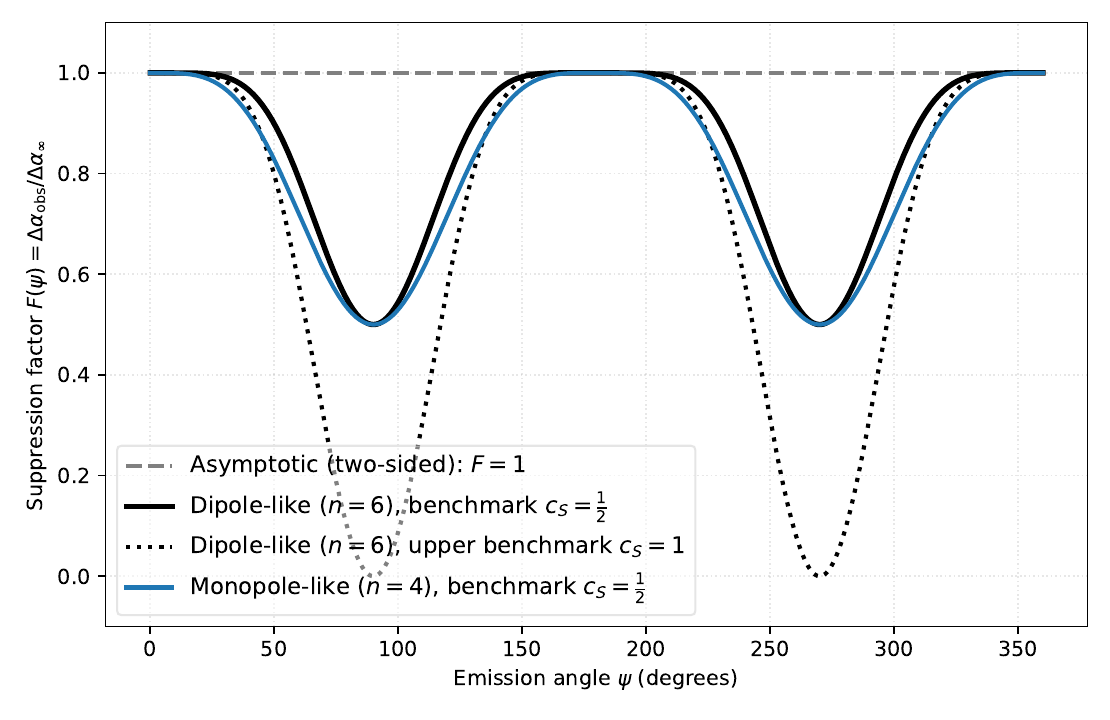}
    \caption{Finite-distance suppression factor for surface emission under benchmark comparisons. We plot the geometric reduction $F(\psi)\equiv \Delta\alpha_{\rm obs}/\Delta\alpha_\infty$ versus emission angle $\psi$. The dashed line shows the asymptotic two-sided scattering normalization ($F=1$). The solid black curve is the dipole-like benchmark $F_{\rm dipole}(\psi)\simeq 1-\tfrac12\sin^{6}\psi$ (Eq. \eqref{V.19}). The dotted black curve shows a more aggressive one-sided benchmark with $c_S=1$, while the blue curve illustrates the monopole-like exponent ($n=4$) at the same reference coefficient $c_S=\tfrac12$, highlighting the sensitivity of the angular dependence to the far-field decay exponent $n$.}
    \label{fig1}
\end{figure}
Figure \ref{fig1} makes explicit what “suppression” means by comparing the one-sided surface-emission geometry to the
two-sided asymptotic scattering normalization ($F=1$). The dipole-like benchmark (solid black) shows that the truncation
is negligible for near-radial emission but becomes rapidly important toward the limb due to the steep $\sin^{6}\psi$
dependence. The dotted curve ($c_S=1$) provides an upper benchmark for how strong the suppression could be under a more
extreme one-sided mapping, while the monopole-like comparison (blue, $n=4$) demonstrates that the exponent controlling
the finite-distance series directly governs how quickly the suppression turns on with emission angle. This overlay therefore
separates the robust geometric trend (strongest effect near the limb) from the model-dependent normalization encoded in
$c_S$ and from the field-structure dependence encoded in $n$.

\subsubsection{Observational consequences for X-ray polarimetry} \label{sec5.3.2}
The implications of Eq. \eqref{V.19} are substantial for the interpretation of data from X-ray polarimetry missions. Standard analyses often assume that the birefringence accumulates over a characteristic length scale $L \sim R_{NS}$, yielding a total phase shift proportional to the full path integral. Our derivation shows that this assumption leads to a systematic overestimation of the NLED coupling strength if the finite-distance truncation is ignored.

More generally, any dipole-like falloff leading to $n=6$ (see Eq. \eqref{V.16}) suggests a one-sided suppression of the form
\begin{equation}
F_{\rm dipole}(\psi)\simeq 1-c_{S}\sin^{6}\psi,
\label{V.20}
\end{equation}
with $c_S=\mathcal{O}(1)$ set by the detailed (generally axisymmetric) geometry and by the mapping between the physical
emission problem and the two-sided scattering normalization used to define $\Delta\alpha_\infty$. Because the field decays rapidly ($B \propto r^{-3}$), the bulk of the birefringence in the asymptotic model is generated in the immediate vicinity of the perigee $r \approx b$. For surface emission, the photon originates at this region of maximum interaction but travels only outward. Consequently, it effectively experiences only half of the lensing potential (or less, depending on $\psi$) compared to a photon passing the star from behind.

\subsubsection{Constraints on physical parameters} \label{sec5.3.3}
The benchmark suppression above indicates that at high emission angles (near-limb emission) the observable birefringence can be
reduced at the order-tens-of-percent level relative to a two-sided full-path integration; in the symmetric one-sided benchmark
($c_S=1/2$) the reduction approaches $\sim 50\%$ as $\psi\to \pi/2$. Such geometric truncation must be accounted for when using
surface-emission polarimetry to constrain the neutron star magnetic field and the effective QED/NLED coupling, otherwise the inferred
parameters will be systematically biased.
Accordingly, any inversion of data should use a model-dependent geometric factor $F(\psi)$:
\begin{equation}
\Delta\alpha_{\rm true}\approx \frac{\Delta\alpha_{\rm measured}}{F(\psi)}\quad\text{(model-dependent)}.\label{4.14}
\end{equation}
By applying this correction, future analyses of magnetar polarization data can decouple the geometric finite-distance effects from genuine deviations in the underlying birefringent vacuum response (e.g. higher-order QED contributions or other birefringent NLED corrections). It ensures that any observed anomaly is attributed correctly to the NLED physics rather than the proximity of the source.

\section{Conclusion} \label{sec6}
In this work, we have developed a rigorous perturbative framework to calculate the finite-distance corrections to vacuum birefringence in strong gravitational and electromagnetic fields. By extending the geometric optical formalism to include the specific case of sources and observers located at finite radial coordinates, we addressed a critical gap in the theoretical modeling of NLED effects near compact objects.

Our analysis began with the linearization of the effective optical metric for weak NLED couplings, isolating the specific contributions of the photon's polarization to the geodesic structure. A key result of our study is the derivation of the generalized differential deflection angle, Eq. \eqref{IV.8}, which explicitly separates the refractive curvature effects from the path-induced geometric shifts. We demonstrated that even in spherically symmetric backgrounds, the splitting of photon trajectories for different polarization modes generates a non-negligible geodesic shift contribution to the observable birefringence.

Applying this formalism to physical scenarios, we computed the explicit finite-distance correction series for Quantum Electrodynamics (Euler-Heisenberg) in a magnetic monopole benchmark and found that the leading finite-distance correction scales as $(b/r)^{4}$. The steep radial dependence signifies that vacuum birefringence is a highly localized phenomenon, with the bulk of the phase shift accumulating in the immediate vicinity of the source. As a consistency check, we also revisited Born-Infeld electrodynamics and confirmed that it is non-birefringent in the present sense: $\Delta\alpha\equiv\alpha_{+}-\alpha_{-}$ vanishes identically \cite{Boillat:1970gw,Obukhov:2002xa}.

Crucially, our application to magnetar physics in Section \ref{sec5.3} revealed that standard asymptotic approximations significantly overestimate the observable polarization rotation for surface emission. Crucially, our magnetar-motivated application in Sec. \ref{sec5.3} shows that standard asymptotic (two-sided) approximations can
overestimate the observable polarization-dependent deflection for surface emission because the physical trajectory is one-sided.
Within a symmetric benchmark mapping this yields a suppression profile $F_{\rm dipole}(\psi)\simeq 1-\tfrac12\sin^{6}\psi$,
indicating order-unity reductions for near-limb rays. Since a realistic dipole magnetosphere is axisymmetric rather than spherically
symmetric, this should be interpreted as an illustrative scaling estimate; a fully axisymmetric extension is left for future work.

Future research should extend this perturbative approach to rotating backgrounds (such as the Kerr spacetime) to evaluate the interplay between frame-dragging and vacuum birefringence. Additionally, investigating the impact of these finite-distance corrections on the polarization signatures of rapid radio bursts (FRBs) originating from magnetar magnetospheres could provide new avenues for testing fundamental physics in extreme environments.

\acknowledgments
R. P. and A. \"O. would like to acknowledge networking support of the COST Action CA21106 - COSMIC WISPers in the Dark Universe: Theory, astrophysics and experiments (CosmicWISPers), the COST Action CA22113 - Fundamental challenges in theoretical physics (THEORY-CHALLENGES), the COST Action CA21136 - Addressing observational tensions in cosmology with systematics and fundamental physics (CosmoVerse), the COST Action CA23130 - Bridging high and low energies in search of quantum gravity (BridgeQG), and the COST Action CA23115 - Relativistic Quantum Information (RQI) funded by COST (European Cooperation in Science and Technology). A. \"O. also thanks to EMU, TUBITAK, ULAKBIM (Turkiye) and SCOAP3 (Switzerland) for their support.

\bibliography{ref}

@article{DeLorenci:2000yh,
    author = "De Lorenci, V. A. and Klippert, R. and Novello, M. and Salim, J. M.",
    title = "{Light propagation in nonlinear electrodynamics}",
    eprint = "gr-qc/0005049",
    archivePrefix = "arXiv",
    doi = "10.1016/S0370-2693(00)00522-0",
    journal = "Phys. Lett. B",
    volume = "482",
    number = "1-3",
    pages = "134--140",
    year = "2000"
}

@article{Novello:1999pg,
    author = "Novello, M. and De Lorenci, V. A. and Salim, J. M. and Klippert, Renato",
    title = "{Geometrical aspects of light propagation in nonlinear electrodynamics}",
    eprint = "gr-qc/9911085",
    archivePrefix = "arXiv",
    reportNumber = "CBPF-NF-050-99",
    doi = "10.1103/PhysRevD.61.045001",
    journal = "Phys. Rev. D",
    volume = "61",
    pages = "045001",
    year = "2000"
}

@article{Novello:2001fv,
    author = "Novello, M. and Salim, J. M. and De Lorenci, V. A. and Elbaz, E.",
    title = "{Nonlinear electrodynamics can generate a closed space - like path for photons}",
    doi = "10.1103/PhysRevD.63.103516",
    journal = "Phys. Rev. D",
    volume = "63",
    pages = "103516",
    year = "2001"
}

@article{Obukhov:2002xa,
    author = "Obukhov, Yuri N. and Rubilar, Guillermo F.",
    title = "{Fresnel analysis of the wave propagation in nonlinear electrodynamics}",
    eprint = "gr-qc/0204028",
    archivePrefix = "arXiv",
    doi = "10.1103/PhysRevD.66.024042",
    journal = "Phys. Rev. D",
    volume = "66",
    pages = "024042",
    year = "2002"
}

@article{Breton:2005ye,
    author = "Breton, Nora",
    title = "{Stability of nonlinear magnetic black holes}",
    eprint = "hep-th/0502217",
    archivePrefix = "arXiv",
    doi = "10.1103/PhysRevD.72.044015",
    journal = "Phys. Rev. D",
    volume = "72",
    pages = "044015",
    year = "2005"
}

@article{Schee:2016mjd,
    author = "Schee, Jan and Stuchl{\'\i}k, Zden",
    title = "{Profiled spectral lines generated by Keplerian discs orbiting in the Bardeen and Ay{\`o}n-Beato{\textendash}Garc{\`\i}a spacetimes}",
    eprint = "1604.00632",
    archivePrefix = "arXiv",
    primaryClass = "gr-qc",
    doi = "10.1088/0264-9381/33/8/085004",
    journal = "Class. Quant. Grav.",
    volume = "33",
    number = "8",
    pages = "085004",
    year = "2016"
}

@article{Stuchlik:2014qja,
    author = "Stuchl{\'\i}k, Zden{\v{e}}k and Schee, Jan",
    title = "{Circular geodesic of Bardeen and Ayon{\textendash}Beato{\textendash}Garcia regular black-hole and no-horizon spacetimes}",
    eprint = "1501.00015",
    archivePrefix = "arXiv",
    primaryClass = "astro-ph.HE",
    doi = "10.1142/S0218271815500200",
    journal = "Int. J. Mod. Phys. D",
    volume = "24",
    number = "02",
    pages = "1550020",
    year = "2014"
}

@article{Schee:2015nua,
    author = "Schee, Jan and Stuchlik, Zdenek",
    title = "{Gravitational lensing and ghost images in the regular Bardeen no-horizon spacetimes}",
    eprint = "1501.00835",
    archivePrefix = "arXiv",
    primaryClass = "astro-ph.HE",
    doi = "10.1088/1475-7516/2015/06/048",
    journal = "JCAP",
    number = "06",
    pages = "048",
    volume = "2015"
}

@article{Schee:2019gki,
    author = "Schee, Jan and Stuchlik, Zdenek",
    title = "{Profiled spectral lines of Keplerian rings orbiting in the regular Bardeen black hole spacetimes}",
    eprint = "1908.07197",
    archivePrefix = "arXiv",
    primaryClass = "gr-qc",
    doi = "10.1140/epjc/s10052-019-7420-1",
    journal = "Eur. Phys. J. C",
    volume = "79",
    number = "12",
    pages = "988",
    year = "2019"
}

@article{Toshmatov:2018tyo,
    author = "Toshmatov, Bobir and Stuchl{\'\i}k, Zden{\v{e}}k and Schee, Jan and Ahmedov, Bobomurat",
    title = "{Electromagnetic perturbations of black holes in general relativity coupled to nonlinear electrodynamics}",
    eprint = "1805.00240",
    archivePrefix = "arXiv",
    primaryClass = "gr-qc",
    doi = "10.1103/PhysRevD.97.084058",
    journal = "Phys. Rev. D",
    volume = "97",
    number = "8",
    pages = "084058",
    year = "2018"
}

@article{Toshmatov:2019gxg,
    author = "Toshmatov, Bobir and Stuchl{\'\i}k, Zden{\v{e}}k and Ahmedov, Bobomurat and Malafarina, Daniele",
    title = "{Relaxations of perturbations of spacetimes in general relativity coupled to nonlinear electrodynamics}",
    eprint = "1903.03778",
    archivePrefix = "arXiv",
    primaryClass = "gr-qc",
    doi = "10.1103/PhysRevD.99.064043",
    journal = "Phys. Rev. D",
    volume = "99",
    number = "6",
    pages = "064043",
    year = "2019"
}

@article{Heisenberg:1936nmg,
    author = "Heisenberg, W. and Euler, H.",
    title = "{Consequences of Dirac's theory of positrons}",
    eprint = "physics/0605038",
    archivePrefix = "arXiv",
    doi = "10.1007/BF01343663",
    journal = "Z. Phys.",
    volume = "98",
    number = "11-12",
    pages = "714--732",
    year = "1936"
}

@article{Schwinger:1951nm,
    author = "Schwinger, Julian S.",
    editor = "Milton, K. A.",
    title = "{On gauge invariance and vacuum polarization}",
    doi = "10.1103/PhysRev.82.664",
    journal = "Phys. Rev.",
    volume = "82",
    pages = "664--679",
    year = "1951"
}

@article{Plebanski:1959ff,
    author = "Plebanski, Jerzy",
    title = "{Electromagnetic Waves in Gravitational Fields}",
    doi = "10.1103/PhysRev.118.1396",
    journal = "Phys. Rev.",
    volume = "118",
    pages = "1396--1408",
    year = "1959"
}

@Article{Einstein_2005,
  author    = {Einstein, A.},
  journal   = {Albert Einstein: Akademie‐Vorträge},
  title     = {Erklärung der Perihelbewegung des Merkur aus der allgemeinen Relativitätstheorie},
  year      = {2005},
  month     = dec,
  pages     = {78--87},
  doi       = {10.1002/3527608958.ch4},
  isbn      = {9783527608959},
  publisher = {Wiley},
}

@article{Dyson:1920cwa,
    author = "Dyson, F. W. and Eddington, A. S. and Davidson, C.",
    title = "{A Determination of the Deflection of Light by the Sun's Gravitational Field, from Observations Made at the Total Eclipse of May 29, 1919}",
    doi = "10.1098/rsta.1920.0009",
    journal = "Phil. Trans. Roy. Soc. Lond. A",
    volume = "220",
    pages = "291--333",
    year = "1920"
}

@article{Born:1934gh,
    author = "Born, M. and Infeld, L.",
    title = "{Foundations of the new field theory}",
    doi = "10.1098/rspa.1934.0059",
    journal = "Proc. Roy. Soc. Lond. A",
    volume = "144",
    number = "852",
    pages = "425--451",
    year = "1934"
}

@article{Born:1933pep,
    author = "Born, M. and Infeld, L.",
    title = "{Foundations of the new field theory}",
    doi = "10.1038/1321004b0",
    journal = "Nature",
    volume = "132",
    number = "3348",
    pages = "1004.1",
    year = "1933"
}

@article{Duncan:1992hi,
    author = "Duncan, Robert C. and Thompson, Christopher",
    title = "{Formation of very strongly magnetized neutron stars - implications for gamma-ray bursts}",
    doi = "10.1086/186413",
    journal = "Astrophys. J. Lett.",
    volume = "392",
    pages = "L9",
    year = "1992"
}

@article{Kaspi:2017fwg,
    author = "Kaspi, Victoria M. and Beloborodov, Andrei",
    title = "{Magnetars}",
    eprint = "1703.00068",
    archivePrefix = "arXiv",
    primaryClass = "astro-ph.HE",
    doi = "10.1146/annurev-astro-081915-023329",
    journal = "Ann. Rev. Astron. Astrophys.",
    volume = "55",
    pages = "261--301",
    year = "2017"
}

@article{Adler:1971wn,
    author = "Adler, Stephen L.",
    title = "{Photon splitting and photon dispersion in a strong magnetic field}",
    doi = "10.1016/0003-4916(71)90154-0",
    journal = "Annals Phys.",
    volume = "67",
    pages = "599--647",
    year = "1971"
}

@article{Gibbons:2008rj,
    author = "Gibbons, G. W. and Werner, M. C.",
    title = "{Applications of the Gauss-Bonnet theorem to gravitational lensing}",
    eprint = "0807.0854",
    archivePrefix = "arXiv",
    primaryClass = "gr-qc",
    doi = "10.1088/0264-9381/25/23/235009",
    journal = "Class. Quant. Grav.",
    volume = "25",
    pages = "235009",
    year = "2008"
}

@article{Ishihara:2016vdc,
    author = "Ishihara, Asahi and Suzuki, Yusuke and Ono, Toshiaki and Kitamura, Takao and Asada, Hideki",
    title = "{Gravitational bending angle of light for finite distance and the Gauss-Bonnet theorem}",
    eprint = "1604.08308",
    archivePrefix = "arXiv",
    primaryClass = "gr-qc",
    doi = "10.1103/PhysRevD.94.084015",
    journal = "Phys. Rev. D",
    volume = "94",
    number = "8",
    pages = "084015",
    year = "2016"
}

@book{Weisskopf_2016,
  author    = {Weisskopf, Martin C. and others},
  title     = {The Imaging X-ray Polarimetry Explorer (IXPE)},
  year      = {2016},
  issn      = {0277-786X},
  month     = jul,
  pages     = {990517},
  volume    = {9905},
  booktitle = {Space Telescopes and Instrumentation 2016: Ultraviolet to Gamma Ray},
  doi       = {10.1117/12.2235240},
  editor    = {den Herder, Jan-Willem A. and Takahashi, Tadayuki and Bautz, Marshall},
  publisher = {SPIE},
}

@article{eXTP:2018anb,
    author = "Zhang, Shuang-Nan and others",
    collaboration = "eXTP",
    title = "{The enhanced X-ray Timing and Polarimetry mission{\textemdash}eXTP}",
    eprint = "1812.04020",
    archivePrefix = "arXiv",
    primaryClass = "astro-ph.IM",
    doi = "10.1007/s11433-018-9309-2",
    journal = "Sci. China Phys. Mech. Astron.",
    volume = "62",
    number = "2",
    pages = "29502",
    year = "2019"
}

@article{Bronnikov:2000vy,
    author = "Bronnikov, Kirill A.",
    title = "{Regular magnetic black holes and monopoles from nonlinear electrodynamics}",
    eprint = "gr-qc/0006014",
    archivePrefix = "arXiv",
    doi = "10.1103/PhysRevD.63.044005",
    journal = "Phys. Rev. D",
    volume = "63",
    pages = "044005",
    year = "2001"
}

@book{Birula_1986,
  author    = {Iwo Bialynicki-Birula},
  title     = {Nonlinear Electrodynamics: Variations on a Theme by Born and Infeld},
  year      = {1983},
  pages     = {31-48},
  booktitle = {Quantum Theory of Particles and Fields: Birthday Volume Dedicated to Jan Łopuszański},
  editor    = {Bernard Jancewicz and Jerzy Lukierski},
  publisher = {World Scientific},
}

@article{Rindler:2007zz,
    author = "Rindler, Wolfgang and Ishak, Mustapha",
    title = "{Contribution of the cosmological constant to the relativistic bending of light revisited}",
    eprint = "0709.2948",
    archivePrefix = "arXiv",
    primaryClass = "astro-ph",
    doi = "10.1103/PhysRevD.76.043006",
    journal = "Phys. Rev. D",
    volume = "76",
    pages = "043006",
    year = "2007"
}

@article{Ishihara:2016sfv,
    author = "Ishihara, Asahi and Suzuki, Yusuke and Ono, Toshiaki and Asada, Hideki",
    title = "{Finite-distance corrections to the gravitational bending angle of light in the strong deflection limit}",
    eprint = "1612.04044",
    archivePrefix = "arXiv",
    primaryClass = "gr-qc",
    doi = "10.1103/PhysRevD.95.044017",
    journal = "Phys. Rev. D",
    volume = "95",
    number = "4",
    pages = "044017",
    year = "2017"
}

@article{Verbin:2024ewl,
    author = {Verbin, Yosef and Pulice, Beyhan and {\"O}vg{\"u}n, Ali and Huang, Hyat},
    title = "{New black hole solutions of second and first order formulations of nonlinear electrodynamics}",
    eprint = "2412.20989",
    archivePrefix = "arXiv",
    primaryClass = "gr-qc",
    doi = "10.1103/PhysRevD.111.084061",
    journal = "Phys. Rev. D",
    volume = "111",
    number = "8",
    pages = "084061",
    year = "2025"
}

@article{Okyay:2021nnh,
    author = {Okyay, Mert and {\"O}vg{\"u}n, Ali},
    title = "{Nonlinear electrodynamics effects on the black hole shadow, deflection angle, quasinormal modes and greybody factors}",
    eprint = "2108.07766",
    archivePrefix = "arXiv",
    primaryClass = "gr-qc",
    doi = "10.1088/1475-7516/2022/01/009",
    journal = "JCAP",
    volume = "01",
    number = "01",
    pages = "009",
    year = "2022"
}

@article{Brihaye:2021ich,
    author = "Brihaye, Y. and Verbin, Y.",
    title = "{Scalarized dyonic black holes in vector-tensor Horndeski gravity}",
    eprint = "2105.11402",
    archivePrefix = "arXiv",
    primaryClass = "gr-qc",
    doi = "10.1103/PhysRevD.104.024047",
    journal = "Phys. Rev. D",
    volume = "104",
    number = "2",
    pages = "024047",
    year = "2021"
}

@article{Sorokin:2021tge,
    author = "Sorokin, Dmitri P.",
    title = "{Introductory Notes on Non-linear Electrodynamics and its Applications}",
    eprint = "2112.12118",
    archivePrefix = "arXiv",
    primaryClass = "hep-th",
    doi = "10.1002/prop.202200092",
    journal = "Fortsch. Phys.",
    volume = "70",
    number = "7-8",
    pages = "2200092",
    year = "2022"
}

@article{Bialynicka-Birula:1970nlh,
    author = "Bialynicka-Birula, Z. and Bialynicki-Birula, I.",
    title = "{Nonlinear effects in Quantum Electrodynamics. Photon propagation and photon splitting in an external field}",
    doi = "10.1103/PhysRevD.2.2341",
    journal = "Phys. Rev. D",
    volume = "2",
    pages = "2341--2345",
    year = "1970"
}

@article{deOliveira:1994in,
    author = "de Oliveira, H. P.",
    title = "{Nonlinear charged black holes}",
    doi = "10.1088/0264-9381/11/6/012",
    journal = "Class. Quant. Grav.",
    volume = "11",
    pages = "1469--1482",
    year = "1994"
}

@article{Pellicer:1969cf,
    author = "Pellicer, R. and Torrence, R. J.",
    title = "{Nonlinear electrodynamics and general relativity}",
    doi = "10.1063/1.1665019",
    journal = "J. Math. Phys.",
    volume = "10",
    pages = "1718--1723",
    year = "1969"
}

@article{Yajima:2000kw,
    author = "Yajima, Hiroki and Tamaki, Takashi",
    title = "{Black hole solutions in Euler-Heisenberg theory}",
    eprint = "gr-qc/0005016",
    archivePrefix = "arXiv",
    doi = "10.1103/PhysRevD.63.064007",
    journal = "Phys. Rev. D",
    volume = "63",
    pages = "064007",
    year = "2001"
}

@article{Ruffini:2013hia,
    author = "Ruffini, Remo and Wu, Yuan-Bin and Xue, She-Sheng",
    title = "{Einstein-Euler-Heisenberg Theory and charged black holes}",
    eprint = "1307.4951",
    archivePrefix = "arXiv",
    primaryClass = "hep-th",
    doi = "10.1103/PhysRevD.88.085004",
    journal = "Phys. Rev. D",
    volume = "88",
    pages = "085004",
    year = "2013"
}

@article{Kruglov:2015yua,
    author = "Kruglov, S. I.",
    title = "{Nonlinear electrodynamics and black holes}",
    eprint = "1504.03941",
    archivePrefix = "arXiv",
    primaryClass = "physics.gen-ph",
    doi = "10.1142/S0219887815500735",
    journal = "Int. J. Geom. Meth. Mod. Phys.",
    volume = "12",
    number = "07",
    pages = "1550073",
    year = "2015"
}

@article{Breton:2021mju,
    author = "Bret{\'o}n, Nora and L{\'o}pez, L. A.",
    title = "{Birefringence and quasinormal modes of the Einstein-Euler-Heisenberg black hole}",
    eprint = "2105.12283",
    archivePrefix = "arXiv",
    primaryClass = "gr-qc",
    doi = "10.1103/PhysRevD.104.024064",
    journal = "Phys. Rev. D",
    volume = "104",
    number = "2",
    pages = "024064",
    year = "2021"
}

@article{Ayon-Beato:1998hmi,
    author = "Ayon-Beato, Eloy and Garcia, Alberto",
    title = "{Regular black hole in general relativity coupled to nonlinear electrodynamics}",
    eprint = "gr-qc/9911046",
    archivePrefix = "arXiv",
    doi = "10.1103/PhysRevLett.80.5056",
    journal = "Phys. Rev. Lett.",
    volume = "80",
    pages = "5056--5059",
    year = "1998"
}

@article{Dymnikova:2004zc,
    author = "Dymnikova, Irina",
    title = "{Regular electrically charged structures in nonlinear electrodynamics coupled to general relativity}",
    eprint = "gr-qc/0407072",
    archivePrefix = "arXiv",
    doi = "10.1088/0264-9381/21/18/009",
    journal = "Class. Quant. Grav.",
    volume = "21",
    pages = "4417--4429",
    year = "2004"
}

@article{Balart:2014cga,
    author = "Balart, Leonardo and Vagenas, Elias C.",
    title = "{Regular black holes with a nonlinear electrodynamics source}",
    eprint = "1408.0306",
    archivePrefix = "arXiv",
    primaryClass = "gr-qc",
    doi = "10.1103/PhysRevD.90.124045",
    journal = "Phys. Rev. D",
    volume = "90",
    number = "12",
    pages = "124045",
    year = "2014"
}

@article{Bambi:2013ufa,
    author = "Bambi, Cosimo and Modesto, Leonardo",
    title = "{Rotating regular black holes}",
    eprint = "1302.6075",
    archivePrefix = "arXiv",
    primaryClass = "gr-qc",
    doi = "10.1016/j.physletb.2013.03.025",
    journal = "Phys. Lett. B",
    volume = "721",
    pages = "329--334",
    year = "2013"
}

@article{Li:2024rbw,
    author = {Li, Zhi-Chao and L{\"u}, Hong},
    title = "{Regular electric black holes from Einstein-Maxwell-scalar gravity}",
    eprint = "2407.07952",
    archivePrefix = "arXiv",
    primaryClass = "gr-qc",
    doi = "10.1103/PhysRevD.110.104046",
    journal = "Phys. Rev. D",
    volume = "110",
    number = "10",
    pages = "104046",
    year = "2024"
}

@article{Ovgun:2019wej,
    author = {{\"O}vg{\"u}n, A.},
    title = "{Weak field deflection angle by regular black holes with cosmic strings using the Gauss-Bonnet theorem}",
    eprint = "1902.04411",
    archivePrefix = "arXiv",
    primaryClass = "gr-qc",
    doi = "10.1103/PhysRevD.99.104075",
    journal = "Phys. Rev. D",
    volume = "99",
    number = "10",
    pages = "104075",
    year = "2019"
}

@article{Fan:2016hvf,
    author = "Fan, Zhong-Ying and Wang, Xiaobao",
    title = "{Construction of Regular Black Holes in General Relativity}",
    eprint = "1610.02636",
    archivePrefix = "arXiv",
    primaryClass = "gr-qc",
    doi = "10.1103/PhysRevD.94.124027",
    journal = "Phys. Rev. D",
    volume = "94",
    number = "12",
    pages = "124027",
    year = "2016"
}

@article{Fradkin:1985qd,
    author = "Fradkin, E. S. and Tseytlin, Arkady A.",
    title = "{Nonlinear Electrodynamics from Quantized Strings}",
    reportNumber = "LEBEDEV-85-193",
    doi = "10.1016/0370-2693(85)90205-9",
    journal = "Phys. Lett. B",
    volume = "163",
    pages = "123--130",
    year = "1985"
}

@article{Vagnozzi:2022moj,
    author = "Vagnozzi, Sunny and others",
    title = "{Horizon-scale tests of gravity theories and fundamental physics from the Event Horizon Telescope image of Sagittarius A}",
    eprint = "2205.07787",
    archivePrefix = "arXiv",
    primaryClass = "gr-qc",
    reportNumber = "UCI-HEP-TR-2022-07",
    doi = "10.1088/1361-6382/acd97b",
    journal = "Class. Quant. Grav.",
    volume = "40",
    number = "16",
    pages = "165007",
    year = "2023"
}

@article{Allahyari:2019jqz,
    author = "Allahyari, Alireza and Khodadi, Mohsen and Vagnozzi, Sunny and Mota, David F.",
    title = "{Magnetically charged black holes from non-linear electrodynamics and the Event Horizon Telescope}",
    eprint = "1912.08231",
    archivePrefix = "arXiv",
    primaryClass = "gr-qc",
    doi = "10.1088/1475-7516/2020/02/003",
    journal = "JCAP",
    number = "02",
    pages = "003",
    volume = "2020"
}

@article{Pantig:2022gih,
    author = {Pantig, Reggie C. and Mastrototaro, Leonardo and Lambiase, Gaetano and {\"O}vg{\"u}n, Ali},
    title = "{Shadow, lensing, quasinormal modes, greybody bounds and neutrino propagation by dyonic ModMax black holes}",
    eprint = "2208.06664",
    archivePrefix = "arXiv",
    primaryClass = "gr-qc",
    doi = "10.1140/epjc/s10052-022-11125-y",
    journal = "Eur. Phys. J. C",
    volume = "82",
    number = "12",
    pages = "1155",
    year = "2022"
}

@article{Lambiase:2005gt,
    author = "Lambiase, G. and Papini, G. and Punzi, Raffaele and Scarpetta, G.",
    title = "{Neutrino optics and oscillations in gravitational fields}",
    eprint = "gr-qc/0503027",
    archivePrefix = "arXiv",
    doi = "10.1103/PhysRevD.71.073011",
    journal = "Phys. Rev. D",
    volume = "71",
    pages = "073011",
    year = "2005"
}

@article{Lambiase:2024lvo,
    author = {Lambiase, Gaetano and Gogoi, Dhruba Jyoti and Pantig, Reggie C. and {\"O}vg{\"u}n, Ali},
    title = "{Shadow and quasinormal modes of the rotating Einstein{\textendash}Euler{\textendash}Heisenberg black holes}",
    eprint = "2406.18300",
    archivePrefix = "arXiv",
    primaryClass = "gr-qc",
    doi = "10.1016/j.dark.2025.101886",
    journal = "Phys. Dark Univ.",
    volume = "48",
    pages = "101886",
    year = "2025"
}

@article{Tseytlin:1995uq,
    author = "Tseytlin, Arkady A.",
    title = "{On singularities of spherically symmetric backgrounds in string theory}",
    eprint = "hep-th/9509050",
    archivePrefix = "arXiv",
    reportNumber = "IMPERIAL-TP-94-95-54",
    doi = "10.1016/0370-2693(95)01228-7",
    journal = "Phys. Lett. B",
    volume = "363",
    pages = "223--229",
    year = "1995"
}

@article{Toshmatov:2021fgm,
    author = "Toshmatov, Bobir and Ahmedov, Bobomurat and Malafarina, Daniele",
    title = "{Can a light ray distinguish charge of a black hole in nonlinear electrodynamics?}",
    eprint = "2101.05496",
    archivePrefix = "arXiv",
    primaryClass = "gr-qc",
    doi = "10.1103/PhysRevD.103.024026",
    journal = "Phys. Rev. D",
    volume = "103",
    number = "2",
    pages = "024026",
    year = "2021"
}

@article{Eiroa:2005ag,
    author = "Eiroa, Ernesto F.",
    title = "{Gravitational lensing by Einstein-Born-Infeld black holes}",
    eprint = "gr-qc/0511065",
    archivePrefix = "arXiv",
    doi = "10.1103/PhysRevD.73.043002",
    journal = "Phys. Rev. D",
    volume = "73",
    pages = "043002",
    year = "2006"
}

@article{Kuang:2022ojj,
    author = "Kuang, Xiao-Mei and Tang, Zi-Yu and Wang, Bin and Wang, Anzhong",
    title = "{Constraining a modified gravity theory in strong gravitational lensing and black hole shadow observations}",
    eprint = "2206.05878",
    archivePrefix = "arXiv",
    primaryClass = "gr-qc",
    doi = "10.1103/PhysRevD.106.064012",
    journal = "Phys. Rev. D",
    volume = "106",
    number = "6",
    pages = "064012",
    year = "2022"
}

@article{Virbhadra:2008ws,
    author = "Virbhadra, K. S.",
    title = "{Relativistic images of Schwarzschild black hole lensing}",
    eprint = "0810.2109",
    archivePrefix = "arXiv",
    primaryClass = "gr-qc",
    doi = "10.1103/PhysRevD.79.083004",
    journal = "Phys. Rev. D",
    volume = "79",
    pages = "083004",
    year = "2009"
}

@article{Virbhadra:1999nm,
    author = "Virbhadra, K. S. and Ellis, George F. R.",
    title = "{Schwarzschild black hole lensing}",
    eprint = "astro-ph/9904193",
    archivePrefix = "arXiv",
    doi = "10.1103/PhysRevD.62.084003",
    journal = "Phys. Rev. D",
    volume = "62",
    pages = "084003",
    year = "2000"
}

@article{Abdujabbarov:2017pfw,
    author = "Abdujabbarov, Ahmadjon and Ahmedov, Bobomurat and Dadhich, Naresh and Atamurotov, Farruh",
    title = "{Optical properties of a braneworld black hole: Gravitational lensing and retrolensing}",
    doi = "10.1103/PhysRevD.96.084017",
    journal = "Phys. Rev. D",
    volume = "96",
    number = "8",
    pages = "084017",
    year = "2017"
}

@article{Atamurotov:2021hoq,
    author = "Atamurotov, Farruh and Abdujabbarov, Ahmadjon and Han, Wen-Biao",
    title = "{Effect of plasma on gravitational lensing by a Schwarzschild black hole immersed in perfect fluid dark matter}",
    doi = "10.1103/PhysRevD.104.084015",
    journal = "Phys. Rev. D",
    volume = "104",
    number = "8",
    pages = "084015",
    year = "2021"
}

@article{Atamurotov:2021qds,
    author = "Atamurotov, Farruh and Abdujabbarov, Ahmadjon and Rayimbaev, Javlon",
    title = "{Weak gravitational lensing Schwarzschild-MOG black hole in plasma}",
    doi = "10.1140/epjc/s10052-021-08919-x",
    journal = "Eur. Phys. J. C",
    volume = "81",
    number = "2",
    pages = "118",
    year = "2021"
}

@article{Turakhonov:2025ojy,
    author = "Turakhonov, Ziyodulla and Atamurotov, Farruh and Ghosh, Sushant G. and Abdujabbarov, Ahmadjon",
    title = "{Probing effects of plasma on shadow and weak gravitational lensing by regular black holes in asymptotically safe gravity}",
    doi = "10.1016/j.dark.2025.101880",
    journal = "Phys. Dark Univ.",
    volume = "48",
    pages = "101880",
    year = "2025"
}

@article{Kumar:2020sag,
    author = "Kumar, Rahul and Islam, Shafqat Ul and Ghosh, Sushant G.",
    title = "{Gravitational lensing by charged black hole in regularized $4D$ Einstein{\textendash}Gauss{\textendash}Bonnet gravity}",
    eprint = "2004.12970",
    archivePrefix = "arXiv",
    primaryClass = "gr-qc",
    doi = "10.1140/epjc/s10052-020-08606-3",
    journal = "Eur. Phys. J. C",
    volume = "80",
    number = "12",
    pages = "1128",
    year = "2020"
}

@article{KumarWalia:2024yxn,
    author = "Kumar Walia, Rahul",
    title = "{Exploring nonlinear electrodynamics theories: Shadows of regular black holes and horizonless ultracompact objects}",
    eprint = "2409.13290",
    archivePrefix = "arXiv",
    primaryClass = "gr-qc",
    doi = "10.1103/PhysRevD.110.064058",
    journal = "Phys. Rev. D",
    volume = "110",
    number = "6",
    pages = "064058",
    year = "2024"
}

@misc{Bokulic:2025brf,
    author = "Bokuli{\'c}, Ana and Juri{\'c}, Tajron and Smoli{\'c}, Ivica",
    title = "{Conundrum of regular black holes with nonlinear electromagnetic fields}",
    eprint = "2510.23711",
    archivePrefix = "arXiv",
    primaryClass = "gr-qc",
    reportNumber = "ZTF-EP-25-07; RBI-ThPhys-2025-41",
    month = "10",
    year = "2025"
}

@misc{Babaei-Aghbolagh:2025tim,
    author = "Babaei-Aghbolagh, H. and Babaei Velni, Komeil and He, Song and Isapour, Fateme",
    title = "{A Unified Causal Framework for Nonlinear Electrodynamics Black Hole from Courant-Hilbert Approach: Thermodynamics and Singularity}",
    eprint = "2511.17407",
    archivePrefix = "arXiv",
    primaryClass = "hep-th",
    month = "11",
    year = "2025"
}

@article{Bokulic:2021xom,
    author = "Bokuli{\'c}, Ana and Juri{\'c}, Tajron and Smoli{\'c}, Ivica",
    title = "{Nonlinear electromagnetic fields in strictly stationary spacetimes}",
    eprint = "2111.10387",
    archivePrefix = "arXiv",
    primaryClass = "gr-qc",
    reportNumber = "ZTF-EP-21-07, RBI-ThPhys-2021-41, ZTF-EP-21-07; RBI-ThPhys-2021-41",
    doi = "10.1103/PhysRevD.105.024067",
    journal = "Phys. Rev. D",
    volume = "105",
    number = "2",
    pages = "024067",
    year = "2022"
}

@article{Bokulic:2022cyk,
    author = "Bokuli{\'c}, Ana and Smoli{\'c}, Ivica and Juri{\'c}, Tajron",
    title = "{Constraints on singularity resolution by nonlinear electrodynamics}",
    eprint = "2206.07064",
    archivePrefix = "arXiv",
    primaryClass = "gr-qc",
    reportNumber = "ZTF-EP-22-02; RBI-ThPhys-2022-23",
    doi = "10.1103/PhysRevD.106.064020",
    journal = "Phys. Rev. D",
    volume = "106",
    number = "6",
    pages = "064020",
    year = "2022"
}

@article{Toshmatov:2018cks,
    author = "Toshmatov, Bobir and Stuchl{\'\i}k, Zden{\v{e}}k and Ahmedov, Bobomurat",
    title = "{Comment on {\textquotedblleft}Construction of regular black holes in general relativity{\textquotedblright}}",
    eprint = "1807.09502",
    archivePrefix = "arXiv",
    primaryClass = "gr-qc",
    doi = "10.1103/PhysRevD.98.028501",
    journal = "Phys. Rev. D",
    volume = "98",
    number = "2",
    pages = "028501",
    year = "2018"
}

@article{Nomura:2020tpc,
    author = "Nomura, Kimihiro and Yoshida, Daisuke and Soda, Jiro",
    title = "{Stability of magnetic black holes in general nonlinear electrodynamics}",
    eprint = "2004.07560",
    archivePrefix = "arXiv",
    primaryClass = "gr-qc",
    reportNumber = "KOBE-COSMO-20-08",
    doi = "10.1103/PhysRevD.101.124026",
    journal = "Phys. Rev. D",
    volume = "101",
    number = "12",
    pages = "124026",
    year = "2020"
}

@article{NiauAkmansoy:2018ilv,
    author = "Niau Akmansoy, P. and Medeiros, L. G.",
    title = "{Constraining nonlinear corrections to Maxwell electrodynamics using $\gamma\gamma$ scattering}",
    eprint = "1809.01296",
    archivePrefix = "arXiv",
    primaryClass = "hep-ph",
    doi = "10.1103/PhysRevD.99.115005",
    journal = "Phys. Rev. D",
    volume = "99",
    number = "11",
    pages = "115005",
    year = "2019"
}

@article{Moreno:2002gg,
    author = "Moreno, Claudia and Sarbach, Olivier",
    title = "{Stability properties of black holes in selfgravitating nonlinear electrodynamics}",
    eprint = "gr-qc/0208090",
    archivePrefix = "arXiv",
    doi = "10.1103/PhysRevD.67.024028",
    journal = "Phys. Rev. D",
    volume = "67",
    pages = "024028",
    year = "2003"
}

@article{Maeda:2021jdc,
    author = "Maeda, Hideki",
    title = "{Quest for realistic non-singular black-hole geometries: regular-center type}",
    eprint = "2107.04791",
    archivePrefix = "arXiv",
    primaryClass = "gr-qc",
    doi = "10.1007/JHEP11(2022)108",
    journal = "JHEP",
    number = "11",
    pages = "108",
    volume = "2022",
    year = "2022"
}

@article{Junior:2025sjr,
    author = "Junior, Ednaldo L. B. and Junior, Jos{\'e} Tarciso S. S. and Lobo, Francisco S. N. and Rodrigues, Manuel E. and da Silva, Lu{\'\i}s F. Dias and Vieira, Henrique A.",
    title = "{Dyonic regular black bounce solutions in general relativity}",
    eprint = "2502.13327",
    archivePrefix = "arXiv",
    primaryClass = "gr-qc",
    doi = "10.1140/epjc/s10052-025-14427-z",
    journal = "Eur. Phys. J. C",
    volume = "85",
    number = "7",
    pages = "724",
    year = "2025"
}

@article{Fouche:2016qqj,
    author = "Fouch{\'e}, M and Battesti, R and Rizzo, C",
    title = "{Limits on nonlinear electrodynamics}",
    eprint = "1605.04102",
    archivePrefix = "arXiv",
    primaryClass = "physics.optics",
    doi = "10.1103/PhysRevD.93.093020",
    journal = "Phys. Rev. D",
    volume = "93",
    number = "9",
    pages = "093020",
    year = "2016",
    note = "[Erratum: Phys.Rev.D 95, 099902 (2017)]"
}

@article{DeFelice:2024seu,
    author = "De Felice, Antonio and Tsujikawa, Shinji",
    title = "{Instability of Nonsingular Black Holes in Nonlinear Electrodynamics}",
    eprint = "2410.00314",
    archivePrefix = "arXiv",
    primaryClass = "gr-qc",
    reportNumber = "YITP-24-124, WUCG-24-09",
    doi = "10.1103/PhysRevLett.134.081401",
    journal = "Phys. Rev. Lett.",
    volume = "134",
    number = "8",
    pages = "081401",
    year = "2025"
}

@article{Cano:2020ezi,
    author = "Cano, Pablo A. and Murcia, {\'A}ngel",
    title = {{Resolution of Reissner-Nordstr{\"o}m singularities by higher-derivative corrections}},
    eprint = "2006.15149",
    archivePrefix = "arXiv",
    primaryClass = "hep-th",
    reportNumber = "IFT-UAM/CSIC-20-99",
    doi = "10.1088/1361-6382/abd923",
    journal = "Class. Quant. Grav.",
    volume = "38",
    number = "7",
    pages = "075014",
    year = "2021"
}

@article{Boillat:1970gw,
    author = "Boillat, G.",
    title = "{Nonlinear electrodynamics - Lagrangians and equations of motion}",
    doi = "10.1063/1.1665231",
    journal = "J. Math. Phys.",
    volume = "11",
    number = "3",
    pages = "941--951",
    year = "1970"
}

@article{Capozziello:2025wwl,
    author = "Capozziello, Salvatore and Battista, Emmanuele and De Bianchi, Silvia",
    title = "{Null geodesics, causal structure, and matter accretion in Lorentzian-Euclidean black holes}",
    eprint = "2507.08431",
    archivePrefix = "arXiv",
    primaryClass = "gr-qc",
    doi = "10.1103/ybjp-8w2w",
    journal = "Phys. Rev. D",
    volume = "112",
    number = "4",
    pages = "044009",
    year = "2025"
}

@article{DeBianchi:2025bgn,
    author = "De Bianchi, Silvia and Capozziello, Salvatore and Battista, Emmanuele",
    title = "{Atemporality from Conservation Laws of Physics in Lorentzian-Euclidean Black Holes}",
    eprint = "2504.17570",
    archivePrefix = "arXiv",
    primaryClass = "gr-qc",
    doi = "10.1007/s10701-025-00848-z",
    journal = "Found. Phys.",
    volume = "55",
    number = "3",
    pages = "36",
    year = "2025"
}

@article{Capozziello:2024ucm,
    author = "Capozziello, Salvatore and De Bianchi, Silvia and Battista, Emmanuele",
    title = "{Avoiding singularities in Lorentzian-Euclidean black holes: The role of~atemporality}",
    eprint = "2404.17267",
    archivePrefix = "arXiv",
    primaryClass = "gr-qc",
    doi = "10.1103/PhysRevD.109.104060",
    journal = "Phys. Rev. D",
    volume = "109",
    number = "10",
    pages = "104060",
    year = "2024"
}

@article{Ono:2017pie,
    author = "Ono, Toshiaki and Ishihara, Asahi and Asada, Hideki",
    title = "{Gravitomagnetic bending angle of light with finite-distance corrections in stationary axisymmetric spacetimes}",
    eprint = "1704.05615",
    archivePrefix = "arXiv",
    primaryClass = "gr-qc",
    doi = "10.1103/PhysRevD.96.104037",
    journal = "Phys. Rev. D",
    volume = "96",
    number = "10",
    pages = "104037",
    year = "2017"
}

@article{Li:2020wvn,
    author = {Li, Zonghai and Zhang, Guodong and \"Ovg\"un, Ali},
    title = "{Circular Orbit of a Particle and Weak Gravitational Lensing}",
    archivePrefix = "arXiv",
    primaryClass = "gr-qc",
    doi = "10.1103/PhysRevD.101.124058",
    journal = "Phys. Rev. D",
    volume = "101",
    number = "12",
    pages = "124058",
    year = "2020"
}

@article{Crisnejo:2018ppm,
    author = "Crisnejo, Gabriel and Gallo, Emanuel and Rogers, Adam",
    title = "{Finite distance corrections to the light deflection in a gravitational field with a plasma medium}",
    eprint = "1807.00724",
    archivePrefix = "arXiv",
    primaryClass = "gr-qc",
    doi = "10.1103/PhysRevD.99.124001",
    journal = "Phys. Rev. D",
    volume = "99",
    number = "12",
    pages = "124001",
    year = "2019"
}

@article{Takahashi:2023eli,
    author = "Takahashi, Kaisei and Kudo, Ryuya and Takizawa, Keita and Asada, Hideki",
    title = "{Equivalence between definitions of the gravitational deflection angle of light for a stationary spacetime}",
    eprint = "2310.00884",
    archivePrefix = "arXiv",
    primaryClass = "gr-qc",
    doi = "10.1103/PhysRevD.108.124011",
    journal = "Phys. Rev. D",
    volume = "108",
    number = "12",
    pages = "124011",
    year = "2023"
}

@article{Ono:2019hkw,
    author = "Ono, Toshiaki and Asada, Hideki",
    title = "{The effects of finite distance on the gravitational deflection angle of light}",
    eprint = "1906.02414",
    archivePrefix = "arXiv",
    primaryClass = "gr-qc",
    doi = "10.3390/universe5110218",
    journal = "Universe",
    volume = "5",
    number = "11",
    pages = "218",
    year = "2019"
}

@article{Fu:2021akc,
    author = "Fu, Qi-Ming and Zhao, Li and Liu, Yu-Xiao",
    title = "{Weak deflection angle by electrically and magnetically charged black holes from nonlinear electrodynamics}",
    eprint = "2101.08409",
    archivePrefix = "arXiv",
    primaryClass = "gr-qc",
    doi = "10.1103/PhysRevD.104.024033",
    journal = "Phys. Rev. D",
    volume = "104",
    number = "2",
    pages = "024033",
    year = "2021"
}

@article{Amaro:2022del,
    author = "Amaro, Daniel and Mac{\'\i}as, Alfredo",
    title = "{Exact lens equation for the Einstein-Euler-Heisenberg static black hole}",
    doi = "10.1103/PhysRevD.106.064010",
    journal = "Phys. Rev. D",
    volume = "106",
    number = "6",
    pages = "064010",
    year = "2022"
}

@article{Chen:2023trn,
    author = "Chen, Yiqian and Wang, Peng and Wu, Houwen and Yang, Haitang",
    title = "{Gravitational lensing by Born-Infeld naked singularities}",
    eprint = "2305.17411",
    archivePrefix = "arXiv",
    primaryClass = "gr-qc",
    doi = "10.1103/PhysRevD.109.084014",
    journal = "Phys. Rev. D",
    volume = "109",
    number = "8",
    pages = "084014",
    year = "2024"
}

@article{Guzman-Herrera:2024fkg,
    author = "Guzman-Herrera, Elda and Montiel, Ariadna and Breton, Nora",
    title = "{Comparative of light propagation in Born-Infeld, Euler-Heisenberg and ModMax nonlinear electrodynamics}",
    eprint = "2407.21326",
    archivePrefix = "arXiv",
    primaryClass = "gr-qc",
    doi = "10.1088/1475-7516/2024/11/002",
    journal = "JCAP",
    volume = "11",
    pages = "002",
    year = "2024"
}

@article{Kim:2021grj,
    author = "Kim, Jin Young",
    title = "{Deflection of light by a Coulomb charge in Born{\textendash}Infeld electrodynamics}",
    eprint = "2104.06246",
    archivePrefix = "arXiv",
    primaryClass = "physics.gen-ph",
    doi = "10.1140/epjc/s10052-021-09291-6",
    journal = "Eur. Phys. J. C",
    volume = "81",
    number = "6",
    pages = "508",
    year = "2021"
}

@article{Kim:2024fle,
    author = "Kim, Jin Young",
    title = "{Deflection of light by a compact object with electric charge and magnetic dipole in Einstein{\textendash}Born{\textendash}Infeld gravity}",
    eprint = "2404.14756",
    archivePrefix = "arXiv",
    primaryClass = "gr-qc",
    doi = "10.1140/epjc/s10052-024-13388-z",
    journal = "Eur. Phys. J. C",
    volume = "84",
    number = "10",
    pages = "1070",
    year = "2024"
}

@article{Mignani:2016fwz,
    author = "Mignani, R. P. and Testa, V. and Caniulef, D. Gonzalez and Taverna, R. and Turolla, R. and Zane, S. and Wu, K.",
    title = "{Evidence for vacuum birefringence from the first optical-polarimetry measurement of the isolated neutron star RX J1856.5{\ensuremath{-}}3754}",
    eprint = "1610.08323",
    archivePrefix = "arXiv",
    primaryClass = "astro-ph.HE",
    doi = "10.1093/mnras/stw2798",
    journal = "Mon. Not. Roy. Astron. Soc.",
    volume = "465",
    number = "1",
    pages = "492--500",
    year = "2017"
}

@misc{Taverna:2022jgl,
    author = "Taverna, Roberto and others",
    title = "{Polarized x-rays from a magnetar}",
    eprint = "2205.08898",
    archivePrefix = "arXiv",
    primaryClass = "astro-ph.HE",
    doi = "10.1126/science.add0080",
    month = "5",
    year = "2022"
}

@article{Lai:2022knd,
    author = "Lai, Dong",
    title = "{IXPE detection of polarized X-rays from magnetars and photon mode conversion at QED vacuum resonance}",
    eprint = "2209.13640",
    archivePrefix = "arXiv",
    primaryClass = "astro-ph.HE",
    doi = "10.1073/pnas.2216534120",
    journal = "Proc. Nat. Acad. Sci.",
    volume = "120",
    number = "17",
    pages = "e2216534120",
    year = "2023"
}

@article{Kim:2024npq,
    author = "Kim, Dong-Hoon and Kim, Chul Min and Kim, Sang Pyo",
    title = "{Strong-field QED effects on polarization states in dipole and quadrudipole pulsar emissions}",
    eprint = "2406.05752",
    archivePrefix = "arXiv",
    primaryClass = "astro-ph.HE",
    doi = "10.1140/epjc/s10052-024-13662-0",
    journal = "Eur. Phys. J. C",
    volume = "84",
    number = "12",
    pages = "1322",
    year = "2024"
}

@article{Taverna:2024uop,
    author = "Taverna, Roberto and Turolla, Roberto",
    title = "{X-ray Polarization from Magnetar Sources}",
    eprint = "2402.05622",
    archivePrefix = "arXiv",
    primaryClass = "astro-ph.HE",
    doi = "10.3390/galaxies12010006",
    journal = "Galaxies",
    volume = "12",
    number = "1",
    pages = "6",
    year = "2024"
}

\end{document}